\documentclass[twocolumn,aps,longbibliography,floatfix,showpacs]{revtex4-1}
\usepackage{mathrsfs,bm,times,amsmath}
\usepackage{lipsum}
\usepackage{graphicx}
\makeindex

\begin{document}

\title{
$\bm p$-orbital density wave with $\bm d$ symmetry 
in high-$\bm{T_c}$ cuprate superconductors
predicted 
\\
by the renormalization-group + constrained RPA theory
}

\author{Masahisa Tsuchiizu}
\author{Youichi Yamakawa}
\author{Hiroshi Kontani}
\affiliation{Department of Physics, Nagoya University, Nagoya 464-8602, Japan}

\date{April 25, 2016}

\begin{abstract}
The discovery of the charge-density-wave
formation in the high-$T_c$ cuprate superconductors
has activated intensive theoretical studies for the pseudogap states.
However, 
the microscopic origin  of the charge-density-wave state 
has been unknown so far
since the many-body effects beyond the mean-field-level approximations,
 called the vertex corrections,
are essential.
Toward solving this problem,
we employ the recently developed functional renormalization-group method, by 
which we can calculate the higher-order vertex corrections in a
 systematic and unbiased way 
with high numerical accuracy.
We discover the critical development of the
$p$-orbital-density-wave ($p$-ODW) instability in the
strong-spin-fluctuation region.
The obtained $p$-ODW state possesses the key characteristics of the
charge ordering pattern in Bi- and Y-based superconductors, such as the
wave vector parallel to the nearest Cu-Cu direction, and the $d$-symmetry
form factor with the antiphase correlation between $p_x$ and $p_y$
orbitals in the same unit cell.
In addition,
from the observation of the 
beautiful scaling relation between the spin susceptibility and
the $p$-ODW susceptibility, we conclude that 
the main driving force of the density wave is the Aslamazov-Larkin vertex
 correction that becomes very singular near the magnetic
 quantum-critical point.
\end{abstract}


\maketitle

\section{INTRODUCTION}

\vspace*{-.3cm}

\begin{figure}[t]
\includegraphics[width=7.7cm,bb=0 0 1114 1803]{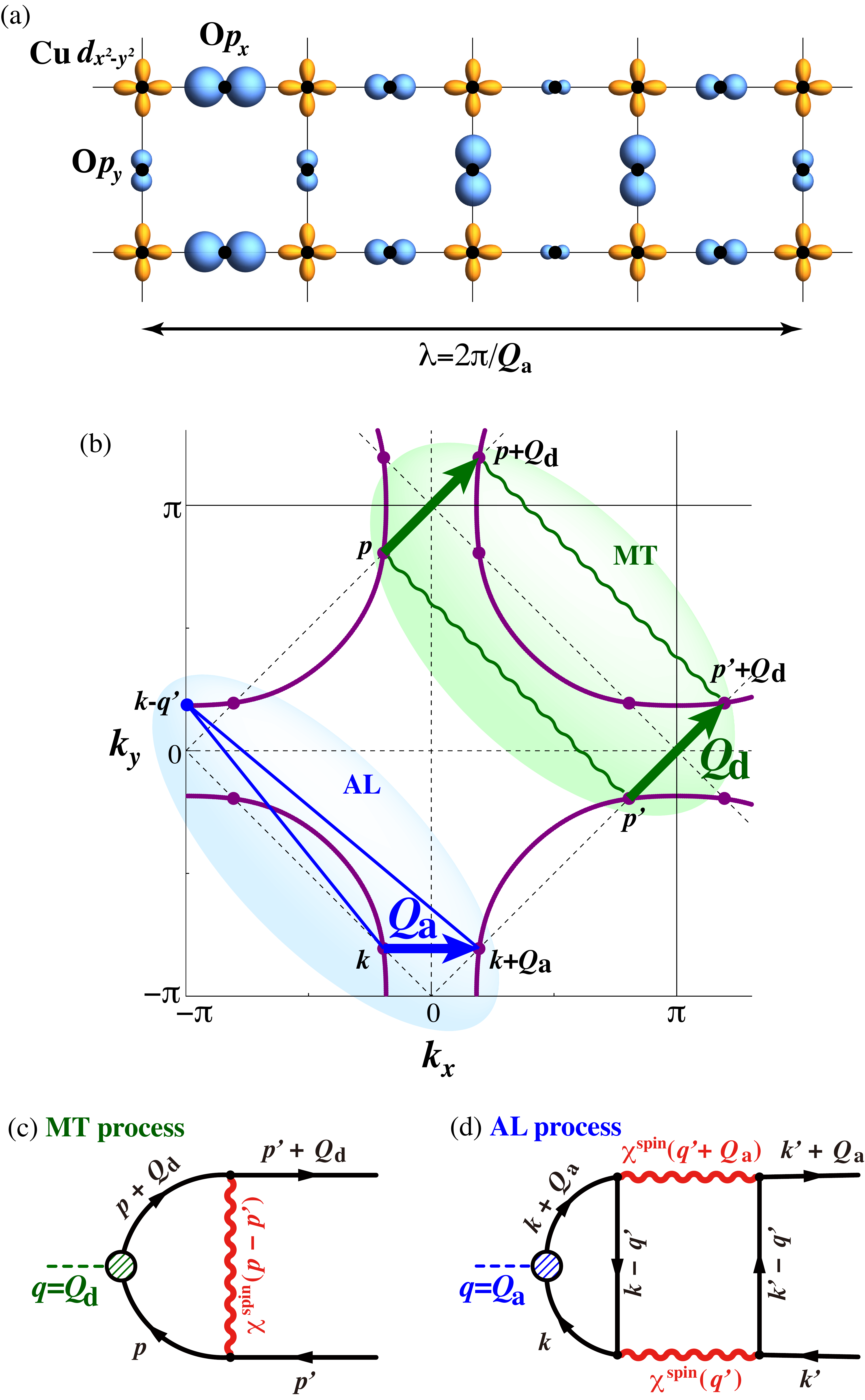}
\caption{
(a) Schematic charge distribution in the 
$d$-symmetry $p$-ODW state with
 the wave vector 
$\bm Q_\mathrm{a} = (0.5\pi, 0)$. 
The size of the orbitals represents the charge density.
(b) The Fermi surface of the cuprate superconductors.
The density-wave instabilities due to the VCs emerge at 
 $\bm Q_\mathrm{d}=(\delta_\mathrm{d},\delta_\mathrm{d})$
and 
 $\bm Q_\mathrm{a}=(\delta_\mathrm{a} ,0)$,
both of which  connect the hot spots.
(c) The Maki-Thompson (MT) process 
\cite{Sachdev:2013bo,%
Holder:2012ks,%
Chowdhury:2014cp%
}
 and 
 (d) the Aslamazov-Larkin (AL) process
 \cite{Yamakawa:2015hba}, which are the first-order and second-order terms
 with respect to the spin  susceptibility, respectively.
The scattering process in the Maki-Thompson VC,  
expressed by $G(\bm p) \, G(\bm p+ \bm Q_\mathrm{d}) \, 
G(\bm p') \, G(\bm p'+\bm Q_\mathrm{d})$,
is shown by green  lines in (b).
Also, the important scattering process in the Aslamazov-Larkin VC, which is expressed by
 $G(\bm k)\, G(\bm k+\bm Q_\mathrm{a})\, G(\bm k-\bm q')$ in the three-point vertex, 
is shown by blue lines
where $\bm q' \approx \bm{Q}_\mathrm{s}$ (the wave vector of the spin fluctuation).
}
\label{fig01}
\end{figure}

Understanding of the
exotic electronic states in underdoped cuprate high-$T_c$
superconductors has been one of the 
greatest challenges in condensed-matter physics.
Especially, the charge-density-wave state observed 
in various compounds is recognized as a key feature of underdoped
cuprate superconductors 
\cite{Bianconi:1996jp%
,Hanaguri:2004hb%
,Daou:2010bo%
,Lawler:2010hp%
,Hucker:2010eq%
,Hucker:2011je%
,Kohsaka:2012hg%
,Ghiringhelli2012sc%
,Chang:2012ib%
,Wu:2013gx%
,Blackburn:2013bs%
,Fujita:2014kg%
,Comin24012014%
,daSilvaNeto:2014tn%
,Tabis:2014kb%
,Wu:2015bt%
,Hamidian:2015eo%
,daSilvaNeto:2015kn%
,Comin20032015%
,Comin:2015ca%
,Comin:2015dn%
}.
The x-ray measurements have succeeded in the direct observation of the
charge-density waves below $\sim$ 200 K
\cite{
Ghiringhelli2012sc,Chang:2012ib%
}.
By means of the resonant x-ray scattering and STM measurements, 
it has now been recognized in many compounds that 
(i) the ordering vector of the charge-density wave is of axial type with
the wave vector 
$\bm Q_\mathrm{a}=(\delta_\mathrm{a},0)$
\cite{%
Ghiringhelli2012sc%
,Chang:2012ib%
,Blackburn:2013bs%
,Comin24012014%
,Fujita:2014kg%
,daSilvaNeto:2014tn%
,Tabis:2014kb%
,Hamidian:2015eo%
,daSilvaNeto:2015kn%
,Comin20032015%
,Comin:2015ca%
,Comin:2015dn%
},
 and 
(ii) the charge modulation emerges mainly on the oxygen $p$ orbital
\cite{%
Fujita:2014kg%
,Hamidian:2015eo%
,Comin:2015ca%
,Comin:2015dn%
}.
In addition, 
(iii) the symmetry of the charge-order pattern is $d$-wave type,
in which the modulations are antiphase 
between $p_x$ and $p_y$  orbitals
in the intra unit cell 
\cite{%
Fujita:2014kg%
,Hamidian:2015eo%
,Comin:2015ca%
,Comin:2015dn%
}.
The charge-modulation pattern
\cite{%
Fujita:2014kg%
,Hamidian:2015eo%
,Comin:2015ca%
,Comin:2015dn%
}
 is depicted in 
Fig.\ \ref{fig01}(a),
which can be regarded as the $d$-symmetry $p$-orbital charge-density-wave
($p$-ODW) state.
The density-wave wave vector $\bm Q_\mathrm{a}$ corresponds to wave vector 
connecting the neighboring ``hot spots''
 shown in Fig.\ \ref{fig01}(b).
The microscopic derivation of the charge-density-wave state 
with the key properties (i)-(iii) based
on the realistic Hubbard model has been desired for years.

Theoretical studies for
the pseudogap phenomena 
have been performed by considering various strong-correlation effects 
\cite{Anderson:1987ii,Varma:1999ca,Emery:1995dr,Lee:2006de,Tremblay:2012im}.
Motivated by the discovery of the density wave below $\sim 200$ K
\cite{%
Ghiringhelli2012sc%
,Chang:2012ib%
,Wu:2013gx%
,Blackburn:2013bs%
,Fujita:2014kg%
,Comin24012014%
,daSilvaNeto:2014tn%
,Tabis:2014kb%
,Wu:2015bt%
,Hamidian:2015eo%
,daSilvaNeto:2015kn%
,Comin20032015%
,Comin:2015ca%
,Comin:2015dn%
},
many scenarios of the
spin-fluctuation-driven nematic order  have been proposed 
in Refs.\ 
\cite{Davis:2013ce,%
Metlitski:2010gf,Metlitski:2010cg,%
Holder:2012ks,
Husemann:2012eb,
Efetov:2013ib,Sachdev:2013bo,Pepin:2014tb,
Allais:2014hm,%
Chowdhury:2014cp,%
Whitsitt:2014vca%
,Berg:2009gt,Lee:2014ka,Fradkin:2015co,Wang:2015uw,Wang:2015iq%
,Wang:2014fr,Tsvelik:2014ce%
},
based on various single-orbital models.
Especially,
 the bond-density-wave (BDW) state 
\cite{Metlitski:2010gf,Metlitski:2010cg,%
Holder:2012ks,
Husemann:2012eb,
Efetov:2013ib,Sachdev:2013bo,Pepin:2014tb,
Allais:2014hm,%
Chowdhury:2014cp,%
Whitsitt:2014vca}, 
the pair-density-wave (PDW) state
\cite{Berg:2009gt,Lee:2014ka,Fradkin:2015co,Wang:2015uw,Wang:2015iq}, and 
the composite charge orders
\cite{Wang:2014fr,Tsvelik:2014ce}
have been studied.
The BDW state is defined as the density wave 
 with $d$-symmetry form factor.
As the driving force of the BDW state,
the authors 
in Refs.\ 
\cite{
Metlitski:2010gf,Metlitski:2010cg,%
Holder:2012ks,%
Sachdev:2013bo,%
Chowdhury:2014cp%
}
had focused on the
Maki-Thompson vertex correction (VC), 
in analogy to the $d$-wave superconductivity driven by the Maki-Thompson VC 
in the Eliashberg theory.
The Maki-Thompson  process for the BDW instability 
is shown in Fig.\ \ref{fig01}(c),
which is the first-order term with respect to the spin susceptibility.
In general, the  
Maki-Thompson VC
gives the density-wave instabilities at
 wave vectors $\bm q=\bm Q_\mathrm{a}$ or $\bm Q_\mathrm{d}$:
The wave vectors
$\bm Q_\mathrm{a}$ and $\bm Q_\mathrm{d}$ 
connect the hot spots on the Fermi surface (FS) shown in Fig.\ \ref{fig01}(b).
The enhancement of the BDW susceptibility
was supported by the renormalization-group (RG)
method 
in the weak-coupling region   \cite{Husemann:2012eb,Whitsitt:2014vca}. 
However,  the predominant wave vector 
is of the diagonal type, 
$\bm Q_\mathrm{d}$,
inconsistently with experimentally-observed axial nematic order.
It was also pointed out that 
  the BDW instability driven by the Maki-Thompson process does not dominate over the 
 superconducting instability \cite{Wang:2014fr,Mishra:2015fb}.
Also,
the PDW state has been considered as the origin of 
 the pseudogap phase \cite{Lee:2014ka,Fradkin:2015co}.
The PDW is formed by the linear combination of the Cooper pairs with
finite momenta. 
The PDW is also induced by the Maki-Thompson process
\cite{Wang:2015uw,Wang:2015iq},
although
 the induced charge modulation has momentum $2\bm Q_\mathrm{a}$
\cite{Fradkin:2015co}.
To realize the density wave with $\bm Q_\mathrm{a}$,
several types of 
the composite charge order parameters have been proposed 
\cite{Wang:2014fr,Tsvelik:2014ce,Wang:2015uw,Wang:2015iq}.
However, 
the $d$ symmetry of its form factor has not been explained.

Quite recently, 
the Aslamazov-Larkin VC
was suggested to be more important near the magnetic 
quantum-critical point \cite{Yamakawa:2015hba}.
The Aslamazov-Larkin VC is the second-order term with respect to the spin
 susceptibility [Fig.\ \ref{fig01}(d)].
The predominant wave vector driven by this VC 
is of the axial type, $\bm{Q}_\mathrm{a}$, 
due to the important scattering process shown in Fig.\ \ref{fig01}(b).
However, 
only the single Aslamazov-Larkin process 
had been studied in Ref.\ \cite{Yamakawa:2015hba}.
In addition, the obtained form factor is given by 
a complex mixture between the $d$- and $p$-orbital charge densities.

In order to settle down the controversy on the driving force of 
the density wave,
we have to employ a sophisticated theoretical method to 
calculate higher-order diagrams,
 including both  Maki-Thompson and Aslamazov-Larkin VCs, 
in a systematic and unbiased way.
For this purpose, the functional RG
method would be the most appropriate theoretical technique.
This method enables us to calculate various types of VCs
up to the parquet-approximation level, 
in which 
the infinite 
 series of the Maki-Thompson and Aslamazov-Larkin VCs are included.
The RG method has been successfully applied 
in the one-dimensional electron systems \cite{Solyom:1979wo,Bourbonnais:2004review}
and has been developed as  
 a powerful method for two-dimensional strongly-correlated electron systems
\cite{Zanchi:1998ua,Halboth:2000vm,Halboth:2000tt,Honerkamp:2001uw,Honerkamp:2001vv,Honerkamp:2005fv,Wang:2009cz,Metzner:2012jv,Wang:2013jg}.

In this paper,
we employ the recently developed improved functional RG
method, called the RG+constrained RPA (cRPA) method  
\cite{Tsuchiizu:2013gu,Tsuchiizu:2015cs},
in order to tackle the unsolved theoretical problem 
on
the charge-density-wave state.
By utilizing this method, we can evaluate the momentum dependence 
 of susceptibilities 
  with high accuracy.
We examine various charge and spin susceptibilities in the 
realistic three-orbital $d$-$p$ Hubbard model 
\cite{Emery:1987dt,Varma:1987cg}. 
We discover that the susceptibility of 
the $d$-symmetry $p$-ODW state 
is critically enhanced at 
 the wave vector $\bm Q_\mathrm{a}$
in the strong-spin-fluctuation region.
The obtained  $p$-ODW state explains satisfactorily 
the experimental key features (i)-(iii) listed above.
Thus, we predict that
the $p$-ODW with $d$ symmetry [Fig.\ \ref{fig01}(a)] is the origin of the
density wave in cuprate superconductors.
The beautiful scaling between the spin and $p$-ODW susceptibilities 
means that
the main driving force of the charge-density instability at  $\bm Q_\mathrm{a}$ 
is the Aslamazov-Larkin VC, 
 which is more singular than the Maki-Thompson VC near
the magnetic quantum-critical point.
Therefore, the $p$-ODW in cuprates originates from the strong
interference between the spin and orbital fluctuations,
which is described microscopically as
the vertex corrections.

This paper is organized as follows.
In Sec.\ \ref{sec:2}, 
we introduce the three-orbital $d$-$p$ model and explain the key idea of 
  the RG+cRPA theory.
In Sec.\ \ref{sec:3}, we show the numerical results on 
  the spin,  $p$-ODW, and $d$-orbital BDW susceptibilities.
In the weak-spin-fluctuation region, 
the diagonal density susceptibilities are
   moderately enhanced 
by the Maki-Thompson VC,
consistently with the previous results
  \cite{Sachdev:2013bo,Chowdhury:2014cp,Husemann:2012eb,Whitsitt:2014vca}. 
In the strong-spin-fluctuation region, in contrast,
 the axial $p$-ODW susceptibility is critically enhanced
 by the Aslamazov-Larkin VC,
dominating over the BDW susceptibility.
In Sec.\ \ref{sec:4},
we show the beautiful scaling relation between the spin and
charge susceptibilities, 
which means that 
the instability at $\bm q=\bm Q_\mathrm{a}$  
   is driven by the Aslamazov-Larkin VC.
We briefly discuss the effect of the Coulomb interaction for $p$ orbital.
Section \ref{sec:5} is devoted to conclusions. 
Details of the
technical calculations are given in Appendixes, where
the comparison between the conventional patch-RG and the
 RG+cRPA theories are also made for the present three-orbital $d$-$p$ model.

\vspace*{-.3cm}

\section{$\bm d$-$\bm p$ Hubbard model and RG+cRPA method}
\label{sec:2}

\vspace*{-.3cm}

\begin{figure*}[t]
\includegraphics[width=12cm,bb=0 0 1362 1103]{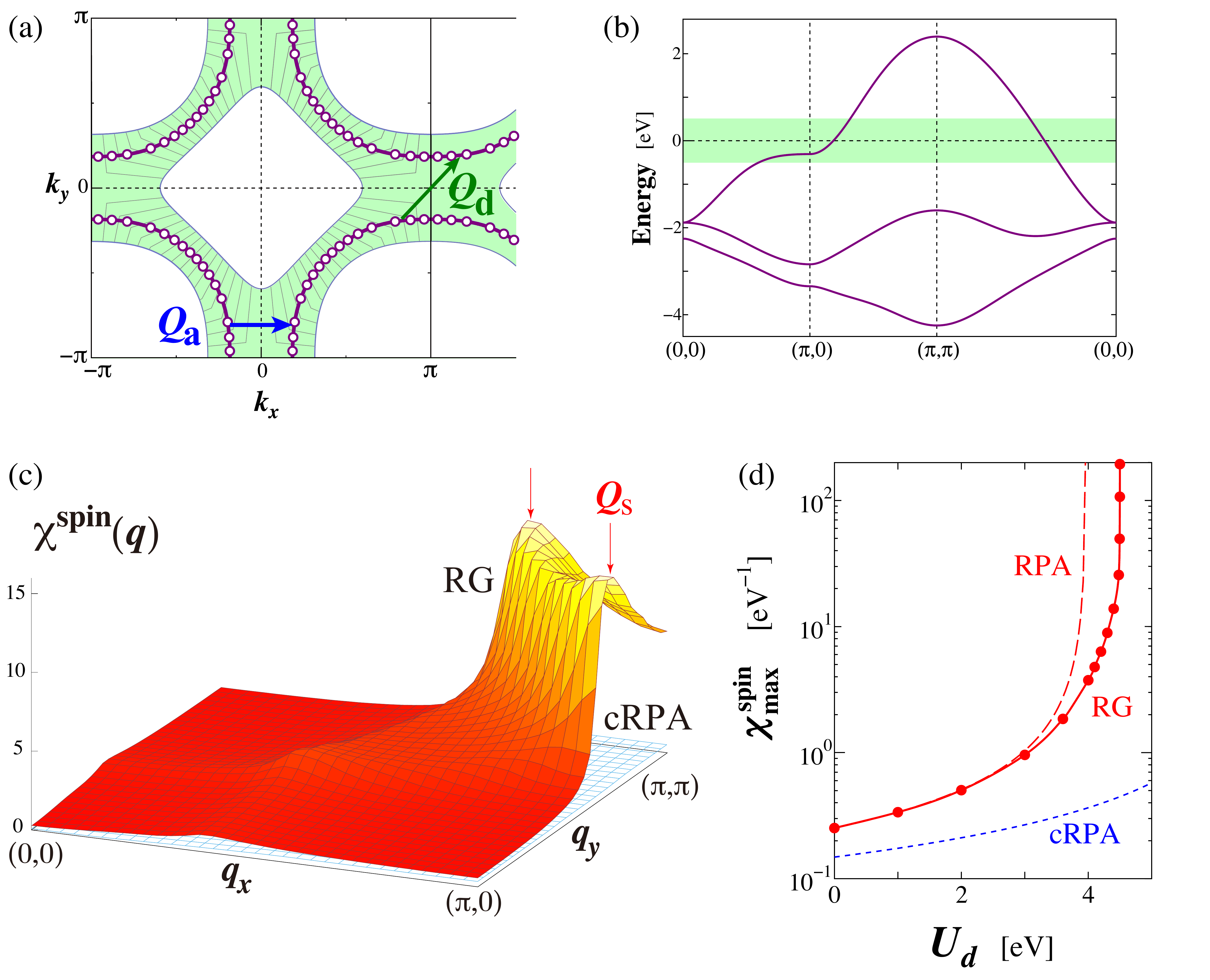}
\caption{
(a) The Fermi surface and
(b) the band structure
in the present three-orbital
$d$-$p$ model.   The low-energy region 
($|E|<0.5$ eV)
are denoted by the shaded areas.  
The $N_p$-patch discretization for 
$N_p =64$ is shown in (a), whereas we set $N_p = 128$ 
in the numerical study of
the present paper.  
(c) The obtained spin susceptibility for the $d$ orbital
$\chi^{\mathrm{spin}}(\bm q)$ for\
$U_d = 4.40$ eV, $T = 20$ meV and $\Lambda_0=0.5$ eV.  
The peak
positions are $\bm q = \bm Q_\mathrm{s} = (\pi-\delta_\mathrm{s},\pi)$ 
and $(\pi,\pi-\delta_\mathrm{s})$ where $\delta_\mathrm{s}\approx 0.31\pi$.  
(d) The  $U_d$ dependencies of 
$\chi^\mathrm{spin}_\mathrm{max} [\equiv \chi^\mathrm{spin}(\bm Q_\mathrm{s})]$ 
given by the RG+cRPA method and by
the RPA.  The initial spin susceptibility given by the cRPA is also
shown.  No ``constrained VC'' discussed in Ref.\ \cite{Tsuchiizu:2013gu}
 is included
in the present RG+cRPA study.
}
\label{fig02}
\end{figure*}

We investigate a standard three-orbital $d$-$p$ Hubbard model
\cite{Emery:1987dt,Varma:1987cg}, 
shown in Fig.\ \ref{fig01}(a),
which has been analyzed for understanding
the charge-density-wave state on the oxygen $p$ orbital
\cite{Bulut:2013bz,Yamakawa:2015hba,Thomson:2015ie}.
 Its Hamiltonian is given by
\begin{equation}
H_{dp}
=
\sum_{\bm k, \sigma}
\bm c_{\bm k, \sigma}^\dagger \,
\hat H_0(\bm k)  \,
\bm c_{\bm k, \sigma}^{}
+U_d\sum_{\bm j}
n_{d, \bm j,\uparrow}
n_{d, \bm j,\downarrow},
\end{equation}
where $\bm c_{\bm k, \sigma}^\dagger=
(d_{\bm k,\sigma}^\dagger, p_{x,\bm k,\sigma}^\dagger, 
 p_{y,\bm k,\sigma}^\dagger)$
is the creation operator for the electron 
on $d_{x^2-y^2}$, $p_x$, and $p_y$ orbitals 
with wave vector $\bm k$
  and spin $\sigma$.
The $d$-electron density operator is
  $n_{d,\bm j,\sigma}=d^\dagger_{\bm j,\sigma}
  d_{\bm j,\sigma}$.
We only consider the on-site Coulomb interaction 
 for the $d$ orbital $U_d$ and the effect of intersite Coulomb
 interaction is briefly discussed later.
For the kinetic term $\hat H_0(\bm k)$,
we use the first-principles hopping integrals for La$_2$CuO$_4$ 
 in Ref.\ \cite{Hansmann:2014ib}
(See the note in Ref.\ \cite{note}).
In addition, we introduce the third-nearest $d$-$d$ hopping $-0.1$ eV 
to make the FS closer to 
Y- and Bi-based cuprates by following Ref.\ \cite{Yamakawa:2015hba}. 
The band filling is set to $n = n_d + n_p = 4.9$, 
corresponding to the hole filling $x = 0.1$. 
The FS and the band structure are shown 
in Figs.\ \ref{fig02}(a) and \ref{fig02}(b), 
respectively.

In underdoped cuprates, the spin, charge, and orbital degrees of
freedoms are strongly coupled by the electron correlation.
This fact had prevented the explanation of the 
experimentally-observed nematic charge order so far.
In order to analyze the competing strong fluctuations
in the two-dimensional systems accurately,
we apply the RG+cRPA method developed in  Refs.\
 \cite{Tsuchiizu:2013gu,Tsuchiizu:2015cs}, 
in which 
the $\bm q$ dependencies of
 susceptibilities can be obtained with high accuracy 
in comparison with the conventional patch-RG method.

In the RG+cRPA method proposed in Ref.\ \cite{Tsuchiizu:2013gu}, 
we set the initial cutoff $\Lambda_0$ smaller than the bandwidth 
[Fig.\ \ref{fig02}(b)].
The scattering processes of electrons having energies
$|E| < \Lambda_0$
 are integrated within the one-loop RG scheme
based on the conventional $N$-patch RG method
\cite{Halboth:2000vm,Halboth:2000tt,Honerkamp:2001uw,Metzner:2012jv}.
The division of lower-energy Brillouin zone ($|E|<\Lambda_0$)
into $N_p$ patches is illustrated in Fig.\ \ref{fig02}(a).
The scattering processes involving higher-energy states
with $|E|>\Lambda_0$
are evaluated by the constrained RPA (cRPA) 
accurately using fine $k$ meshes
and incorporated into
 the initial values. 
This treatment is based on the natural assumption that the vertex 
corrections become significant only in the low-energy regions 
$|E|\ll \Lambda_0$. 
Due to this treatment,
low numerical accuracy for the higher-energy processes 
inherent in the conventional patch-RG method 
is greatly improved in the RG+cRPA method.
Another advantage of the RG+cRPA theory for multi-orbital systems 
is that the inter-band processes 
(so-called van Vleck contributions) can be included.
Therefore, the susceptibilities at low temperatures
 are obtained with high accuracy in the RG+cRPA method.

In the present analysis we set 
$\Lambda_0=0.5$ eV and $N_p=128$.
We have checked the case for $\Lambda_0=1.0$ eV
 and find that 
the results are essentially independent of $\Lambda_0$.
In the present RG+cRPA method, 
the numerical accuracy is 
  sufficiently improved in comparison with the 
conventional patch-RG method.
However,
the main results of this paper are robust and reproduced  
qualitatively
 even in the conventional patch-RG method.
The direct comparison between the numerical results of 
the RG+cRPA and those of 
 the conventional patch-RG method
are made in  Appendix \ref{sec:app-A}.

\vspace*{-.3cm}

\section{Numerical Results}\label{sec:3}

\vspace*{-.3cm}
\subsection{Spin susceptibility}

\vspace*{-.3cm}

First, 
we analyze the spin susceptibility
using the RG+cRPA method, by following the procedure explained in Refs.\ 
\cite{Tsuchiizu:2013gu,Tsuchiizu:2015cs}.
Because of the $d$-orbital Coulomb interaction $U_d$, 
the spin fluctuations develop only on the $d$ orbital. 
The $d$-orbital spin susceptibility per spin is given as
\begin{equation}
\chi^\mathrm{spin}(\bm q) 
=
\frac{1}{2}
\int_0^{1/T} d\tau 
\left\langle  S_{d}(\bm q,\tau) \, S_{d}(-\bm q,0)
\right\rangle ,
\label{eq:d-orb-spin}
\end{equation}
where $S_{d}(\bm q,\tau)$ is the spin operator for the $d$ orbital.
The momentum dependence of the obtained $\chi^\mathrm{spin}(\bm q)$ is shown in 
Fig.\ \ref{fig02}(c).
The strong spin fluctuations 
are realized at the
incommensurate wave vectors
$\bm Q_\mathrm{s}
 =(\pi-\delta_\mathrm{s},\pi)$ and $(\pi,\pi-\delta_\mathrm{s})$.
The obtained $\bm q$ dependence of $\chi^\mathrm{spin}(\bm q)$ is 
consistent with the neutron measurements.
As increasing $U_d$, 
the maximum of the spin susceptibility, $\chi^\mathrm{spin}_\mathrm{max}$,
 develops 
monotonically and diverges 
at $U_d=U_d^\mathrm{cr}(\approx 4.5 \mbox{ eV})$,
as shown in Fig.\ \ref{fig02}(d).
The value of $U_d^\mathrm{cr}$ will increase
to the first-principles value $U_d\approx 8$ eV by including the
spin fluctuation-induced 
self-energy \cite{Kontani:2008fd}.

As seen from Figs.\ \ref{fig02}(c) and \ref{fig02}(d), 
the contributions from the cRPA are small
for $\Lambda_0=0.5$ eV but quite important 
(especially for the four-point vertex
 \cite{Tsuchiizu:2013gu,Tsuchiizu:2015cs})
 in order to derive reliable results.
In order to verify the validity of the RG+cRPA theory, 
we analyzed the same $d$-$p$ model by using
the conventional patch-RG method established in literature 
 \cite{Halboth:2000vm,Halboth:2000tt,Honerkamp:2001uw,Metzner:2012jv}
in Appendix \ref{sec:app-A}:
It is confirmed that 
the essential results given by the RG+cRPA method presented in the
main text are \textit{qualitatively} reproduced by the conventional patch-RG
method, whereas the numerical accuracy is well improved in the RG+cRPA method.

\vspace*{-.3cm}
\subsection{$\bm p$-ODW susceptibility and $\bm d$-orbital BDW susceptibility}

\vspace*{-.3cm}

In contrast to spin susceptibility, 
any charge susceptibilities are not enhanced by $U_d$ in the
mean-field-level approximations.
Nonetheless, we reveal that 
the strong $p$-ODW instability emerges 
in the present RG analysis,
thanks to the VCs that are dropped in the RPA.
Since $U_p=0$,
the effective interaction on $p$ orbitals 
that causes the $p$-ODW instability  is originated from the 
many-body effects on the $d$ orbital.
The $p$-ODW susceptibility per spin is defined as
\begin{eqnarray}
\chi^{p\mbox{-}\mathrm{orb}}_{\alpha\beta}(\bm q) 
&=&
\frac{1}{2}
\int_0^{1/T} d\tau 
\left\langle  n_{\alpha}(\bm q,\tau) \, 
n_{\beta}(-\bm q,0)
\right\rangle,
\label{eq:chi-def}
\end{eqnarray}
where $\alpha,\beta=x,y$ represent 
 the $p_x,p_y$ orbitals.
According to the experimental analyses 
\cite{Fujita:2014kg,Hamidian:2015eo}, 
we introduce the susceptibilities 
  for the $p$-ODW with $d$ and $s'$ symmetries:
\begin{eqnarray}
\chi^{p\mbox{-}\mathrm{orb}}_{d}(\bm q) 
&\equiv&
   \chi^{p\mbox{-}\mathrm{orb}}_{xx}(\bm q) 
+  \chi^{p\mbox{-}\mathrm{orb}}_{yy}(\bm q) 
- 2 \chi^{p\mbox{-}\mathrm{orb}}_{xy}(\bm q) 
,
\\
\chi^{p\mbox{-}\mathrm{orb}}_{s'}(\bm q) 
&\equiv&
   \chi^{p\mbox{-}\mathrm{orb}}_{xx}(\bm q) 
+  \chi^{p\mbox{-}\mathrm{orb}}_{yy}(\bm q) 
+ 2 \chi^{p\mbox{-}\mathrm{orb}}_{xy}(\bm q) .
\end{eqnarray}
The susceptibility $\chi^{p\mbox{-}\mathrm{orb}}_{d}(\bm q) $ measures the 
development of the antiphase correlation between $p_x$ and $p_y$ orbitals 
in the same unit cell. 
Aside from the conventional charge/orbital orders, 
it was pointed out in literature 
\cite{Metlitski:2010gf,%
  Metlitski:2010cg,Sachdev:2013bo,%
  Holder:2012ks,Thomson:2015ie,Allais:2014hm}
that the bond-density order with $d$-wave form factor is expected to develop in
the Hubbard model.
Therefore, we calculate the $d$-orbital BDW susceptibility:
\begin{eqnarray}
\chi^\mathrm{BDW}(\bm q) 
&=&
\frac{1}{2}
\int_0^{1/T} d\tau 
\left\langle  B(\bm q,\tau) \, 
B(-\bm q,0)
\right\rangle,
\\
B(\bm q)
&=&
\sum_{\bm k,\sigma} f(\bm k +\bm q/2) \, d_{\bm k,\sigma}^\dagger 
d_{\bm k+\bm q,\sigma}^{},
\end{eqnarray}
where 
$f(\bm k)$ is the $d$-wave form factor:
$f(\bm k)=\cos (k_x) - \cos (k_y)$.
The order parameter $\langle B(\bm q)\rangle\ne 0$ represents the
bond-ordered state,
which is equivalent to the modulation of the hopping integrals
\cite{Allais:2014hm}.
In the case of $\bm q=\bm 0$, the susceptibility $\chi^{\mathrm{BDW}}(\bm q=\bm 0)$ 
  measures the nematic or Pomeranchuk instability 
  \cite{Halboth:2000vm,Halboth:2000tt,Honerkamp:2001uw,Metzner:2012jv}.
We also evaluate the conventional  $d$-orbital charge susceptibility
\begin{eqnarray}
\chi^{d\mbox{-}\mathrm{orb}}(\bm q) 
=
\frac{1}{2}
\int_0^{1/T} d\tau 
\left\langle  n_d(\bm q,\tau) \, 
n_d(-\bm q,0)
\right\rangle,
\label{eq:d-orb-charge}
\end{eqnarray}
where $n_d(\bm q,\tau)$ is the density operator for the $d$ orbital.
Although the upper limit of 
$\chi^{d\mbox{-}\mathrm{orb}}(\bm q) $
 is $1/U_d$ in the present Hubbard model,
$\chi^{d\mbox{-}\mathrm{orb}}(\bm q) $ in the one-loop RG may
 diverge unphysically
  if we set $U_d$ too large.
Thus, we always verify the non-singular behavior of
$\chi^{d\mbox{-}\mathrm{orb}}(\bm q)$ 
 to ensure
that the adopted parameter value is within a
  valid range.
We applied the finite-temperature RG formalism based 
 on the sharp band-width cutoff 
and solved the RG equations down to $\Lambda_l=0$.
In the actual calculation of the RG equations, 
we introduce a lower-energy cutoff 
 $\Lambda_l=\pi T$
for the four-point vertex function in order to obtain stable numerical
results \cite{Husemann:2012eb}.
The physical meaning of this lower-energy cutoff is, for example, 
 the suppression of the Cooper channel due to the impurity scattering
or the magnetic field.
Experimentally, the charge-density-wave state  is strongly stabilized 
by applying the magnetic field beyond 15 T
\cite{Gerber:2015gx,Chang:2015wy}.
This fact means that
the Cooper instability is less important 
for the density-wave-formation mechanism.

\begin{figure*}[t]
\includegraphics[width=13cm,bb=0 0 736 959]{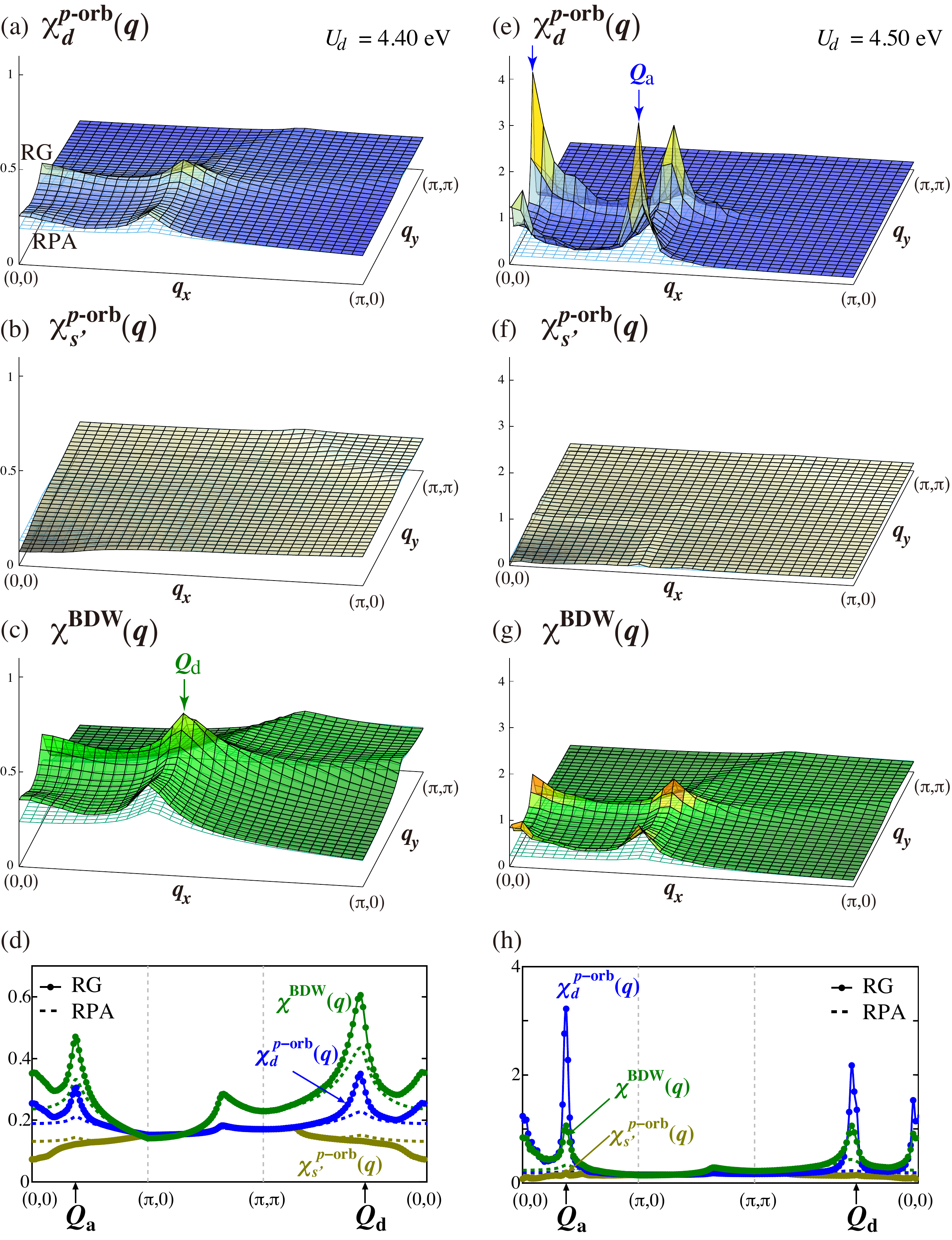}
\caption{
The $p$-ODW [$\chi^{p\mbox{-}\mathrm{orb}}_{d}(\bm q)$ and
 $\chi^{p\mbox{-}\mathrm{orb}}_{s'}(\bm q)$]
and  $d$-orbital BDW [$\chi^\mathrm{BDW}(\bm q)$] susceptibilities in the
three-orbital Hubbard model with $U_d = 4.40$ eV in (a)-(d), 
  and $U_d=4.50$ eV in (e)-(h).
The temperature is $T = 20$ meV.  
The maximum of the spin susceptibility 
is  $\chi^\mathrm{spin}_\mathrm{max}\approx 12$ eV$^{-1}$ 
  ($200$ eV$^{-1}$) for 
  $U_d=4.40$ eV ($4.50$ eV). 
 The RPA results are also shown for comparison.
The axial wave vector is $\bm Q_\mathrm{a}\approx  (0.37\pi, 0)$ 
and the diagonal wave vector is
$\bm Q_\mathrm{d}\approx  (0.41\pi,0.41\pi)$.  
Both peak positions $\bm Q_\mathrm{a}$ and $\bm Q_\mathrm{d}$ 
correspond to the
wave vector connecting the hot spots shown in Fig.\ \ref{fig01}(b).  
The wave vectors
$\bm Q_\mathrm{a}$ and $\bm Q_\mathrm{d}$ 
obtained from the RG+cRPA method change with carrier doping,
by following the change in the 
vectors connecting the hot spots.
}
\label{fig03}
\end{figure*}

In Fig.\ \ref{fig03}, we present the obtained $p$-ODW 
susceptibilities
with $d$ and $s'$ symmetries, 
together with the 
$d$-orbital BDW 
susceptibility for $U_d=4.40$ eV in (a)-(d), and $U_d=4.50$ eV in (e)-(h).
In the case of $U_d=4.40$ eV,
in which the spin susceptibility is moderate 
 ($\chi^\mathrm{spin}_\mathrm{max}\approx 12 $ eV$^{-1}$),
both  $\chi^{p\mbox{-}\mathrm{orb}}_{d}$ and $\chi^{\mathrm{BDW}}$ are
enlarged
compared to the RPA results
as shown in Figs.\ \ref{fig03}(a) and \ref{fig03}(c), whereas 
$\chi^{p\mbox{-}\mathrm{orb}}_{s'}$ in Fig.\ \ref{fig03}(b) is not
enhanced at all.
Therefore,
the $p$-ODW and BDW susceptibilities 
are moderately enhanced by 
the VCs that are neglected in the RPA.
However, the highest peaks of both 
$\chi^{p\mbox{-}\mathrm{orb}}_{d}$ and $\chi^{\mathrm{BDW}}$
are located at 
$\bm q=\bm Q_\mathrm{d}$,
inconsistently with the axial 
nematic density wave in cuprates.

In the case of $U_d=4.50$ eV,
in which the spin susceptibility 
is large  ($\chi^\mathrm{spin}_\mathrm{max}\approx 200 $ eV$^{-1}$),
both the $p$-ODW and $d$-orbital BDW susceptibilities shown in
Figs.\ \ref{fig03}(e) and \ref{fig03}(g)  possess large sharp peaks at 
$\bm q= \bm Q_\mathrm{a}$ and $\bm Q_\mathrm{d}$,
which originate from the VCs generated in the renormalization procedure.
The most divergent density-wave susceptibility is 
$\chi^{p\mbox{-}\mathrm{orb}}_{d}(\bm q)$ at $\bm q=\bm Q_\mathrm{a}$.
In contrast, $\chi^{p\mbox{-}\mathrm{orb}}_{s'}(\bm q)$ is seldom enhanced.
Therefore, the axial $p$-ODW with $d$ symmetry shown in Fig.\ \ref{fig01}(a) is realized.
We have verified that 
$|\bm Q_\mathrm{a}|=\delta_\mathrm{a}$ increases with hole doping
in the present RG study, consistently with experiments in the Y-, 
Bi-, and Hg-based compounds.
According to neutron inelastic scattering studies, 
$\chi^\mathrm{spin}_{\rm max}$ 
is as large as $\sim 500$ eV$^{-1}$ in the slightly underdoped YBCO
\cite{Stock:2004gu}.
Therefore, the nematic density wave in cuprates is realized in the strong-spin
fluctuation region experimentally.

We note that
both 
$\chi^{p\mbox{-}\mathrm{orb}}_{d}$ and $\chi^{\mathrm{BDW}}$ have
sub-dominant  broad peaks at $\bm q=\bm 0$, indicating
that ``the $\bm q=\bm 0$ Pomeranchuk instabilities'' are also enhanced
in the present model.
This instability had been reported in the previous theoretical studies
 \cite{Husemann:2012eb,Caprara:2005gv,Tassini:2005gu,Devereaux:2007bb},
and also observed experimentally in cuprates as the enhancement of the 
  $B_{1g}$ channel Raman response 
 \cite{Tassini:2005gu,Devereaux:2007bb,Caprara:2005gv}.
The temperature-flow RG scheme \cite{Honerkamp:2001vv}
would also be useful
for the study of the Pomeranchuk instability
\cite{Halboth:2000vm,Honerkamp:2005fv}.

\begin{figure*}[t]
\includegraphics[width=16cm,bb=0 0 586 242]{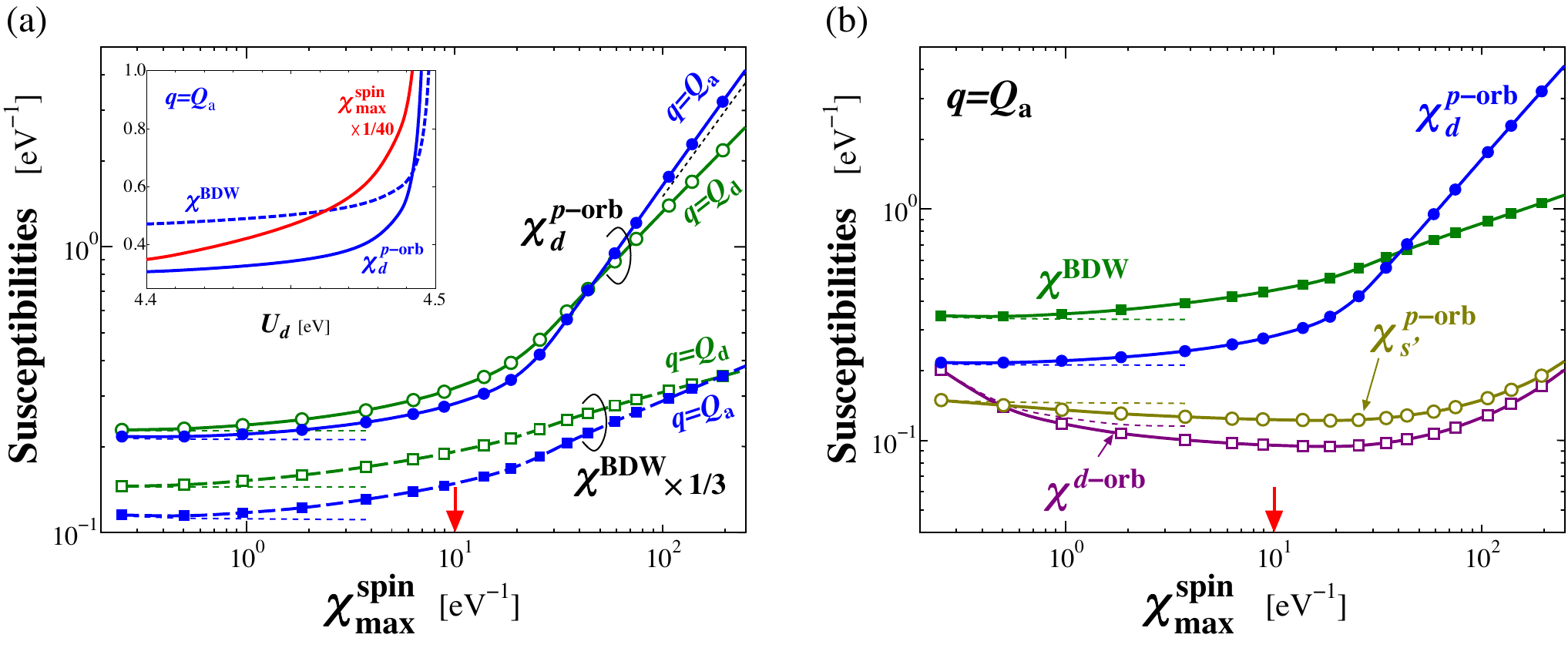}
\caption{
(a)
The $p$-ODW and $d$-orbital BDW susceptibilities,
$\chi^{p\mbox{-}\mathrm{orb}}_{d}(\bm q)$ 
and $\chi^\mathrm{BDW}(\bm q)$,
at peak positions
as functions of $\chi^\mathrm{spin}_\mathrm{max}$.
The short-dashed lines
represent the RPA results, which are almost independent of $U_d$.  
$\chi^{p\mbox{-}\mathrm{orb}}_{d}(\bm q)$ starts to increase when
$\chi^\mathrm{spin}_\mathrm{max}$ exceeds 10 eV$^{-1}$ 
(shown by the arrow).
The corresponding Stoner enhancement is
 $\chi^\mathrm{spin}_\mathrm{max}/\chi_0 \approx 40$.
The gradient of the dotted line is 1, so
a beautiful scaling relation $\chi^{p\mbox{-}\mathrm{orb}}_{d} \propto
\chi^\mathrm{spin}_\mathrm{max}$ is satisfied
in the strong-spin-fluctuation region ($\chi^{\mathrm{spin}}_\mathrm{max}>10$).
The inset shows the $U_d$ dependencies of
$\chi^{p\mbox{-}\mathrm{orb}}_{d}(\bm q)$ and $\chi^\mathrm{BDW}(\bm q)$
at $\bm q= \bm Q_\mathrm{a}$.
(b)
The $d$-symmetry and $s'$-symmetry $p$-ODW susceptibilities, 
$\chi^{p\mbox{-}\mathrm{orb}}_{d}(\bm q)$ and 
$\chi^{p\mbox{-}\mathrm{orb}}_{s'}(\bm q)$,
as well as the $d$-orbital BDW and charge
susceptibilities,
$\chi^{\mathrm{BDW}}(\bm q)$ and 
$\chi^{d\mbox{-}\mathrm{orb}}(\bm q)$,
 at $\bm q=\bm Q_\mathrm{a}$,
as functions of $\chi^\mathrm{spin}_\mathrm{max}$. 
In the strong-spin-fluctuation region,
 $\chi^{p\mbox{-}\mathrm{orb}}_{d}(\bm q)$
  exhibits a critical enhancement, whereas
 the $d$-orbital charge susceptibility 
$\chi^{d\mbox{-}\mathrm{orb}}(\bm q)$ remains
  small. 
Thus, the enhancement of $\chi^{p\mbox{-}\mathrm{orb}}_{d}(\bm Q_\mathrm{a})$ 
is obtained 
  within the reliable parameter range of the RG method.
}
\label{fig04}
\end{figure*}

\section{Discussions}\label{sec:4}

\vspace*{-.3cm}
\subsection{Scaling relation between $\bm p$-ODW susceptibility  \\
 and spin susceptibility}\label{subsec:4a}

\vspace*{-.3cm}

In the inset of Fig.\ \ref{fig04}(a), we show the $U_d$ dependencies 
of $p$-ODW and $d$-orbital BDW susceptibilities at $\bm q=\bm Q_\mathrm{a}$.
The $p$-ODW susceptibility exceeds the BDW one
with increasing $U_d$.
 In the main figure of Fig.\ \ref{fig04}(a),
we plot both the $p$-ODW and $d$-orbital BDW susceptibilities 
at $\bm q=\bm Q_\mathrm{a}$ and $\bm Q_\mathrm{d}$ 
as functions of 
$\chi^\mathrm{spin}_\mathrm{max}$,
in order to reveal the correlation between the spin and 
 density susceptibilities.
In the weak-spin-fluctuation region 
($\chi_\mathrm{max}^\mathrm{spin} \lesssim 10$ eV$^{-1}$), 
both the $p$-ODW and $d$-orbital BDW susceptibilities
increase moderately.
In the strong-spin-fluctuation region,
only $\chi_{d}^{p\mbox{-}\mathrm{orb}}(\bm q)$ 
starts to 
increase drastically in proportion to $\chi^\mathrm{spin}_\mathrm{max}$.
In this region, the highest peak of
$\chi_{d}^{p\mbox{-}\mathrm{orb}}(\bm q)$ 
shifts to $\bm q=\bm Q_\mathrm{a}$,
consistently with the experimental wave vector.
The relation $\chi_{d}^{p\mbox{-}\mathrm{orb}}(\bm Q_\mathrm{a}) \gg
\chi^{\mathrm{BDW}}(\bm Q_\mathrm{a})$ is robust against the choice of 
model parameters.

The most important finding in Fig.\ \ref{fig04}(a) is that 
$\chi_{d}^{p\mbox{-}\mathrm{orb}}(\bm q)$ 
at $\bm q=\bm Q_\mathrm{a}$ well scales 
to $\chi^\mathrm{spin}_\mathrm{max}$ in the strong-fluctuation region 
($\chi^\mathrm{spin}_\mathrm{max}\gtrsim 10$ eV$^{-1}$). 
This beautiful scaling relation 
is obtained in the wide range of model parameters.
This fact indicates that the $p$-ODW is driven by the Aslamazov-Larkin VC
that describes the strong interference 
between spin  and orbital fluctuations.
Also,
as shown in Fig.\ \ref{fig04}(b), 
$\chi_{s'}^{p\mbox{-}\mathrm{orb}}(\bm q)$ 
as well as the conventional charge susceptibility for the $d$ orbital 
$\chi^{d\mbox{-}\mathrm{orb}}(\bm q)$  decrease with 
$\chi^\mathrm{spin}_\mathrm{max}$ in the weak-spin-fluctuation region,
whereas they turn to increase slightly in the strong-spin-fluctuation region.
The obtained relation $\chi^{d\mbox{-}\mathrm{orb}}(\bm q)< 1/U_d$ 
supports the reliability 
 of the present RG result even for
 $\chi^\mathrm{spin}_\mathrm{max}\sim 200 $ eV$^{-1}$.
%

\begin{figure*}[t]
\includegraphics[width=15.5cm,bb=0 0 1164 655]{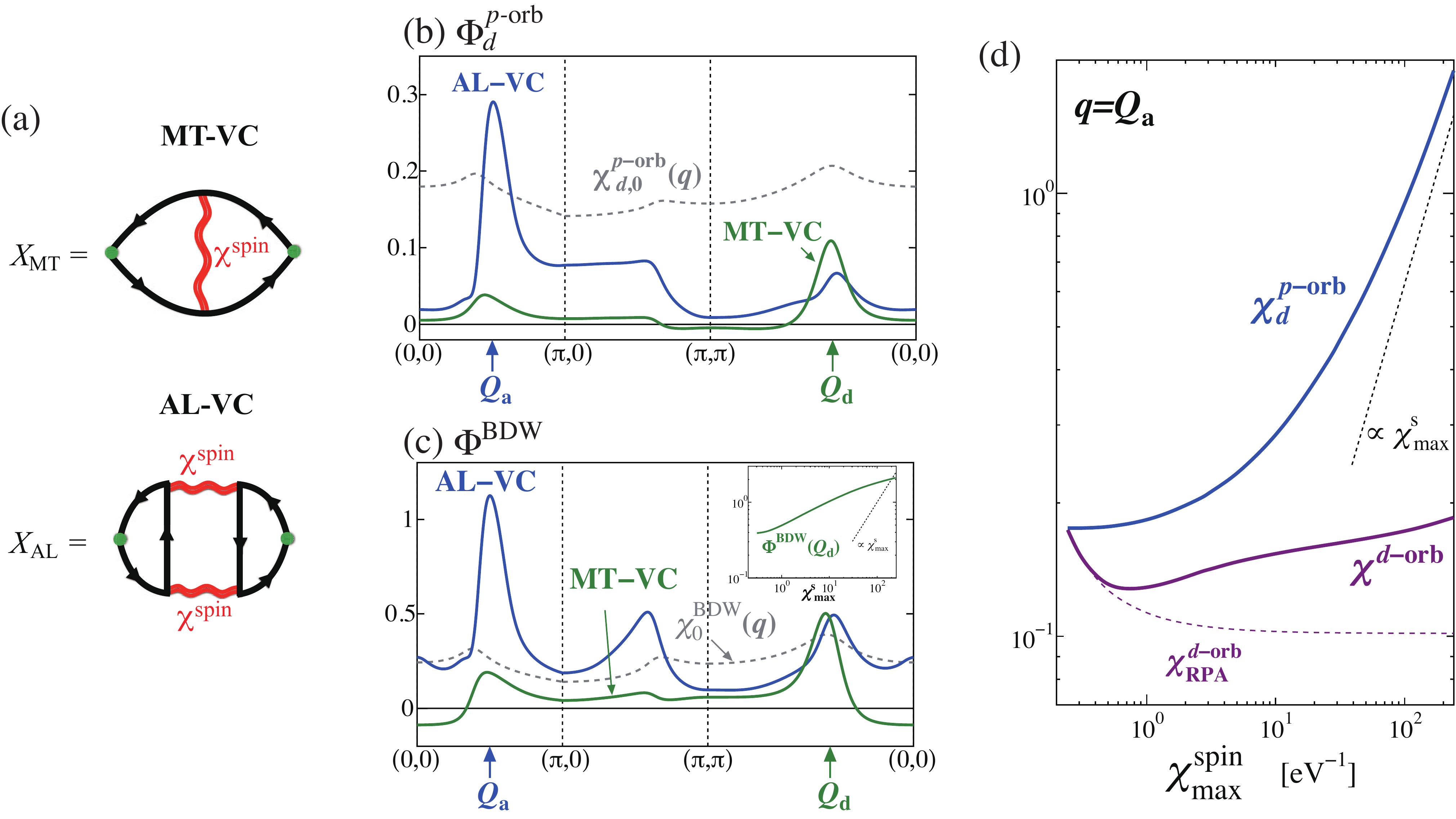}
\caption{
(a) 
The lowest-order 
Maki-Thompson VC
$X_\mathrm{MT}(\bm q)$ and Aslamazov-Larkin VC $X_\mathrm{AL}(\bm q)$.  
Their exact mathematical expressions are
given in Appendix \ref{sec:app-C}.
(b) The momentum dependencies of 
the Maki-Thompson and Aslamazov-Larkin VCs
 for the $p$-ODW susceptibility, at $T = 50$ meV.
The magnitude of $U_d$ is set to satisfy 
the Stoner factor $\alpha_s = 0.99$.
The noninteracting susceptibility 
$\chi^{p\mbox{-}\mathrm{orb}}_{d,0}(\bm q)$ is also shown.  
(c) The Maki-Thompson and Aslamazov-Larkin VCs for the $d$-orbital BDW
susceptibility, 
for $\alpha_s =0.99$ with $T = 50$ meV.  
At $\bm q=\bm Q_\mathrm{d}$,
the Maki-Thompson and Aslamazov-Larkin VCs are comparable.
The inset shows that $\Phi^\mathrm{BDW}(\bm Q_\mathrm{d})$ does not scale to 
$\chi^\mathrm{spin}_\mathrm{max}$. 
(d)
The full susceptibilities $\chi^{p\mbox{-}\mathrm{orb}}_d(\bm Q_\mathrm{a})$ and
$\chi^{d\mbox{-}\mathrm{orb}}(\bm Q_\mathrm{a})$ as functions of
 $\chi^{\mathrm{spin}}_{\mathrm{max}}$.
The scaling relation 
$\chi^{p\mbox{-}\mathrm{orb}}_d(\bm Q_\mathrm{a})\propto \chi^\mathrm{spin}_\mathrm{max}$ 
is well satisfied in the strong-spin-fluctuation region.  
For comparison, the RPA result for $\chi^{d\mbox{-}\mathrm{orb}}(\bm Q_\mathrm{a})$
is also shown.
}
\label{fig05}
\end{figure*}

\vspace*{-.3cm}
\subsection{
Why is the axial $\bm p$-ODW susceptibility enlarged in the
  strong-spin-fluctuation region?
}
\label{subsec:4b}

\vspace*{-.3cm}

In order to understand the physical origin of the $p$-ODW 
and, in addition, in order to confirm the validity of the present RG analysis,
 we also perform the diagrammatic analysis for
the $d$-$p$ Hubbard model.
Hereafter, we demonstrate that the characteristic behaviors of the $p$-ODW and
BDW susceptibilities are well understood by taking 
VCs for the irreducible susceptibilities.
The lowest-order Maki-Thompson  and Aslamazov-Larkin VCs,
$X_\mathrm{MT}(\bm q)$ and
$X_\mathrm{AL}(\bm q)$,
are shown diagrammatically in Fig.\ \ref{fig05}(a).
Their exact expressions are given 
in Appendix \ref{sec:app-C}.
In the strong-spin-fluctuation region,
the Aslamazov-Larkin VC is scaled as 
\cite{Yamakawa:2015hba}
\begin{equation}
X_\mathrm{AL}(\bm Q_\mathrm{a})
\, \propto \,
\sum _{\bm q} 
\chi^\mathrm{spin}(\bm q) \, \chi^\mathrm{spin}(\bm q+\bm Q_\mathrm{a})
\, \propto \, 
\xi^2
\, \propto \,
\chi^\mathrm{spin}_\mathrm{max},
\end{equation}
where $\xi$ is the magnetic correlation length,
while the Maki-Thompson VC is scaled as
\begin{equation}
X_\mathrm{MT}(\bm Q_\mathrm{a,d})\propto \sum _{\bm q} 
 \chi^\mathrm{spin}(\bm q+\bm Q_\mathrm{a,d})
\propto \ln \xi^2
 \propto \ln \chi^\mathrm{spin}_\mathrm{max}.
\end{equation}
Thus, the Aslamazov-Larkin VC is expected to dominate over 
the Maki-Thompson VC in the strong-spin-fluctuation region.

Figure \ref{fig05}(b) shows the $\bm q$ dependencies of the Maki-Thompson and
Aslamazov-Larkin VCs for the $p$-ODW susceptibility, at $U_d=4.06$ eV and $T=50$ meV.
In this case, the system is in the intermediate-spin-fluctuation region
with $\chi^\mathrm{spin}_\mathrm{max}\approx  24 $ eV$^{-1}$.
Here the Aslamazov-Larkin VC possesses the highest peak at 
$\bm q=\bm Q_\mathrm{a}$, 
and the Maki-Thompson VC has the second highest peak at  $\bm q=\bm Q_\mathrm{d}$.
Thus, the irreducible susceptibility 
\begin{equation}
\Phi^{p\mbox{-}\mathrm{orb}}_{d}(\bm q)
\equiv
\chi^{p\mbox{-}\mathrm{orb}}_{d,0}(\bm q)
  + X^{p\mbox{-}\mathrm{orb}}_{d,\mathrm{MT}}(\bm q)
  + X^{p\mbox{-}\mathrm{orb}}_{d,\mathrm{AL}}(\bm q)
\end{equation}
has the largest peak at  $\bm q=\bm Q_\mathrm{a}$ due to the Aslamazov-Larkin VC.

We also show
the Maki-Thompson  and Aslamazov-Larkin VCs 
for the $d$-orbital BDW susceptibility
in Fig.\ \ref{fig05}(c).
Since both VCs have large peaks at 
$\bm q= \bm Q_\mathrm{d}$, the irreducible susceptibility
\begin{equation}
\Phi^{\mathrm{BDW}}(\bm q) 
\equiv
\chi^\mathrm{BDW}_{0}(\bm q)
  + X^\mathrm{BDW}_{\mathrm{MT}}(\bm q)
  + X^\mathrm{BDW}_{\mathrm{AL}}(\bm q)
\end{equation}
takes the highest peak at $\bm q= \bm Q_\mathrm{d}$.
As shown in the inset of Fig.\ \ref{fig05}(c),
$\Phi^{\mathrm{BDW}}(\bm q)$
at  $\bm q= \bm Q_\mathrm{d}$  
does not scale to $\chi_\mathrm{max}^\mathrm{spin}$,
which is consistent with the smallness of $\chi^{\rm BDW}(\bm q)$ 
in Fig.\ \ref{fig04}(a).

Next, we derive the full susceptibilities 
from the irreducible susceptibilities. 
Figure \ \ref{fig05}(d) shows the obtained
$\chi^{p\mbox{-}\mathrm{orb}}_d(\bm q)$ and
$\chi^{d\mbox{-}\mathrm{orb}}(\bm q)$ 
at $\bm q= \bm Q_\mathrm{a}$ 
as functions of $\chi^\mathrm{spin}_\mathrm{max}$
by changing $U_d$.
Since $U_p=0$ in the present model,
 the $p$-ODW susceptibility is well approximated as 
$\chi^{p\mbox{-}\mathrm{orb}}_{d}(\bm q) \approx
\Phi^{p\mbox{-}\mathrm{orb}}_{d}(\bm q)$.
The obtained $\chi^{p\mbox{-}\mathrm{orb}}_{d}(\bm Q_\mathrm{a})$
behaves very similarly to the RG result in Fig.\ \ref{fig04}(a).
Especially, the scaling relation 
$\chi^{p\mbox{-}\mathrm{orb}}_{d}(\bm Q_\mathrm{a}) \propto  
\chi^\mathrm{spin}_\mathrm{max}$ is well reproduced
by the diagrammatic analysis, due to large contribution from the Aslamazov-Larkin VC.
In contrast, the $d$-orbital charge susceptibility
\begin{equation}
\chi^{d\mbox{-}\mathrm{orb}}(\bm q)
=\frac{\Phi^{d\mbox{-}\mathrm{orb}}(\bm q)}
{1+ U_d \,\Phi^{d\mbox{-}\mathrm{orb}}(\bm q)} ,
\end{equation}
at $\bm q=\bm Q_\mathrm{a}$
shows a minimum at finite $\chi^\mathrm{spin}_\mathrm{max}$,
similarly to the RG result in Fig.\ \ref{fig04}(b).
This behavior is also understood analytically: 
In the weak-spin-fluctuation region, in which the VCs are negligible and
therefore
$\Phi^{d\mbox{-}\mathrm{orb}} (\bm q)\approx
\chi^{d\mbox{-}\mathrm{orb}}_0(\bm q)$
is satisfied,
the $d$-orbital susceptibility decreases with $U_d$ in proportion  to
$1/[1+U_d \chi^{d\mbox{-}\mathrm{orb}}_0(\bm q)]$.
In the strong-spin-fluctuation region, 
$\Phi^{d\mbox{-}\mathrm{orb}}(\bm q)$ increases drastically because of the
Aslamazov-Larkin VC, and therefore $\chi^{d\mbox{-}\mathrm{orb}}(\bm q)$ increases 
toward $1/U_d$.

Thus, we revealed that the characteristic  behaviors of the $p$-ODW susceptibility
in the present RG study,
such as 
the peak position at $\bm q=\bm Q_\mathrm{a}$
and the scaling relation 
$\chi^{p\mbox{-}\mathrm{orb}}_d(\bm Q_\mathrm{a}) 
\propto  \chi^\mathrm{spin}_\mathrm{max}$,
 are qualitatively understood by including the 
 Aslamazov-Larkin VC into the RPA.
This result is never trivial in that the higher-order VCs,
unrestricted to the Maki-Thompson  and Aslamazov-Larkin VCs, are systematically produced in the
RG theory.
For example, 
the higher-order 
Maki-Thompson and Aslamazov-Larkin 
processes are included in the RG.
Also, the spin and charge fluctuations and the four-point VCs 
are calculated consistently.
Thus, the dominant role of the lowest-order Aslamazov-Larkin VC shown in Fig.\
\ref{fig05} is 
confirmed in the present RG theory.

Of course, 
the lowest-order Aslamazov-Larkin VC study cannot explain the RG results in many parts.
For example, 
   if  only the lowest-order Aslamazov-Larkin VC is included,
the relation
$\chi^\mathrm{BDW}(\bm Q_\mathrm{a}) \propto 
\chi^\mathrm{spin}_\mathrm{max}$
is realized 
although it remains small in the RG results.
Also,
the $d$-symmetry form factor in the $p$-ODW cannot simply be obtained
by the lowest-order Aslamazov-Larkin VC.
These facts indicate the importance of the higher-order diagrams
   included in the RG.

Note that there is another type of the Aslamazov-Larkin VC
 given by the product of spin 
and charge (or orbital) propagators 
$\sim \chi^\mathrm{spin}(\bm q')\, \chi^\mathrm{charge}(\bm q'')$.
This gives rise to the correction 
 to the spin susceptibility $\chi^\mathrm{spin}$
and is also included in the present RG formalism.
From the data of spin susceptibility [Fig.\ \ref{fig02}(d)],
it can be confirmed that such a VC does not have a strong effect.

\begin{figure}[t]
\includegraphics[width=7cm,bb=0 0 578 459]{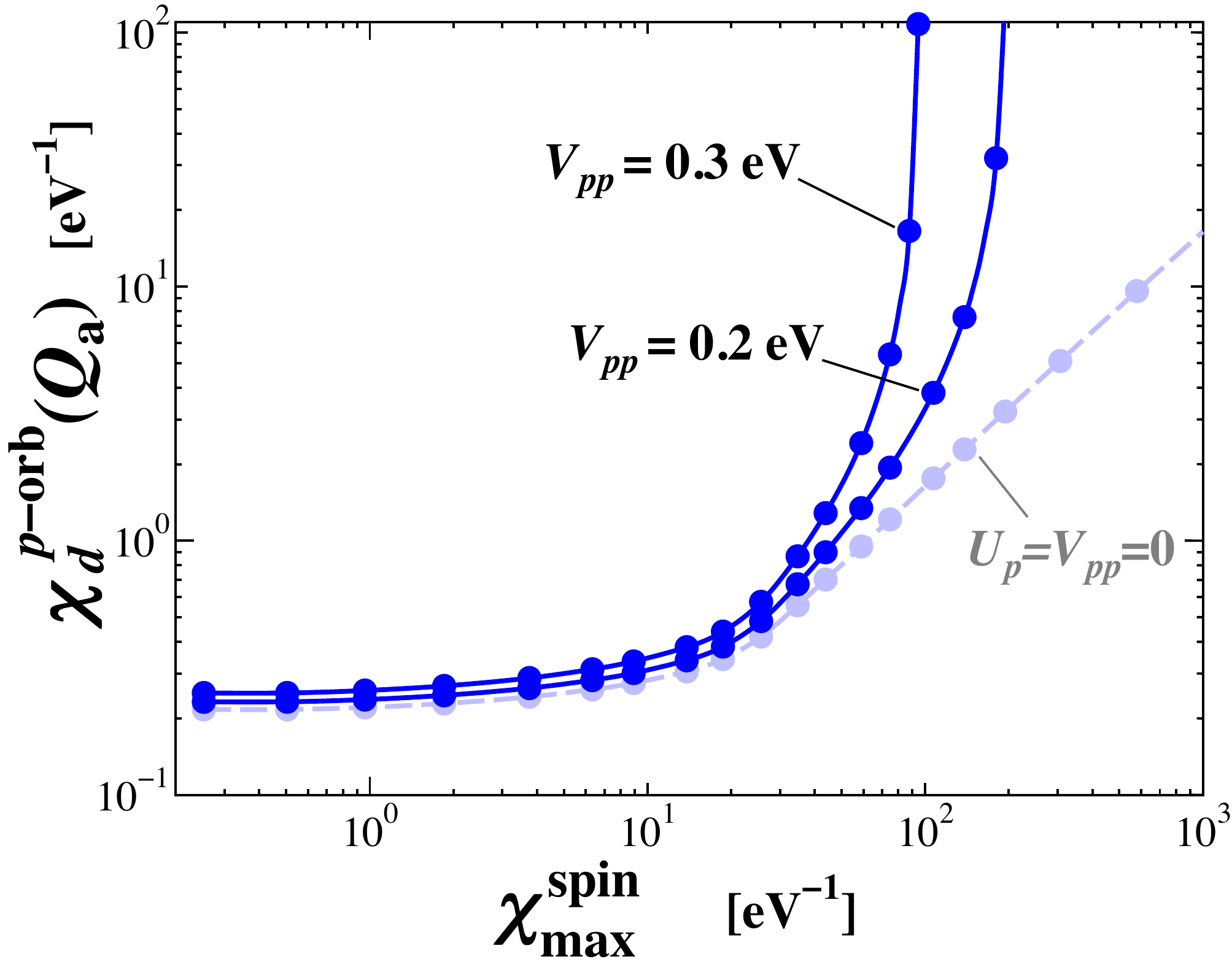}

\caption{
The $p$-ODW susceptibility
$\chi^{p\mbox{-}\mathrm{orb}}_d(\bm Q_\mathrm{a}; U_p,V_{pp})$
 [Eq.\ (\ref{eq:chi-pODW-V})]
as a function of $\chi^{\mathrm{spin}}_\mathrm{max}$
for $U_p=0.7$ eV.
As an approximation, the irreducible susceptibility is  given by 
 the RG result for $U_p=V_{pp}=0$.
}
\label{fig06}
\end{figure}

\vspace*{-.3cm}
\subsection{Effect of the Coulomb interactions for O sites on the 
  $\bm p$-ODW susceptibility}
\label{subsec:4c}

\vspace*{-.3cm}

The scaling relation $\chi^{p\mbox{-}\mathrm{orb}}_d(\bm Q_\mathrm{a}) \propto 
 \chi^\mathrm{spin}_\mathrm{max}$ 
obtained in Fig.\ \ref{fig04} 
indicates that the spin-density-wave order and the $p$-ODW order emerge
 simultaneously.
However,
the $p$-ODW will be realized in the paramagnetic state 
in the presence of  small but finite Coulomb interaction 
between the nearest-neighbor $p$-$p$ orbitals, $V_{pp}$.
The $p$-ODW susceptibility
is given as
\begin{equation}
\hat\chi^{p\mbox{-}\mathrm{orb}}(\bm q; U_p,V_{pp}) 
=\frac{\hat \chi^{p\mbox{-}\mathrm{orb}}(\bm q)}
{1+ \hat V(\bm q) \, \hat \chi^{p\mbox{-}\mathrm{orb}}(\bm q) },
\label{eq:chi-pODW-V}
\end{equation}
where $\hat\chi^{p\mbox{-}\mathrm{orb}}(\bm q)$ 
represents the irreducible susceptibility with respect to 
$U_p$ (the $p$-orbital onsite interaction) and $V_{pp}$.
The non-zero matrix elements of $\hat V(\bm q)$ are
 $V_{xx}= V_{yy}=U_p$ and 
 $V_{xy}= V_{yx}=8V_{pp} \cos (q_x/2) \cos (q_y/2)$.
When $U_p,V_{pp}$ are very small,
$\hat \chi^{p\mbox{-}\mathrm{orb}}(\bm q)$  is safely approximated by 
  the RG susceptibility for $U_p=V_{pp}=0$.
Figure \ref{fig06} shows the obtained 
$\chi^{p\mbox{-}\mathrm{orb}}_{d}(\bm Q_\mathrm{a};U_p,V_{pp})$
 for $V_{pp}=0.2$ and $0.3$ eV with $U_p=0.7$ eV,
which diverges even when $\chi^\mathrm{spin}_\mathrm{max}$ is finite. 
Now, we explain why the tiny
$V_{pp}$ critically enhances the $p$-ODW susceptibility.
Considering the relation 
 $\chi^{p\mbox{-}\mathrm{orb}}_{xx}
  \chi^{p\mbox{-}\mathrm{orb}}_{yy} 
  \approx [\chi^{p\mbox{-}\mathrm{orb}}_{xy}]^2$
 obtained in the present RG at $\bm q=\bm Q_\mathrm{a} $,
the $d$-symmetry $p$-ODW susceptibility 
is approximately given as
\begin{widetext}
\begin{eqnarray}
\chi^{p\mbox{-}\mathrm{orb}}_{d}(\bm Q_\mathrm{a}; U_p, V_{pp})
\approx 
\frac{\chi^{p\mbox{-}\mathrm{orb}}_{d}(\bm Q_\mathrm{a}) }
{1 + 
   U_p \, [\chi^{p\mbox{-}\mathrm{orb}}_{xx}(\bm Q_\mathrm{a}) 
          +\chi^{p\mbox{-}\mathrm{orb}}_{yy}(\bm Q_\mathrm{a}) ]
  +2  V_{xy}(\bm Q_\mathrm{a}) \,
       \chi^{p\mbox{-}\mathrm{orb}}_{xy}(\bm Q_\mathrm{a}) }
,
\end{eqnarray}
\end{widetext}
which is enhanced by $V_{pp}$ since  
$\chi^{p\mbox{-}\mathrm{orb}}_{xy}(\bm Q_\mathrm{a})$ is negative.
The enhancement due to $V_{pp}$ dominates over the 
suppression due to $U_p$, if 
 $U_p\lesssim 8 V_{pp} \cos \delta_\mathrm{a}$.
In real materials, finite electron-phonon coupling 
would also enlarge the $p$-ODW susceptibility 
in the strong-spin-fluctuation region. 
Therefore, the Aslamazov-Larkin VC accounts for the 
$p$-ODW ordering in the pseudogap region of the cuprate superconductors.

\vspace*{-.3cm}

\section{Conclusions}\label{sec:5}

\vspace*{-.3cm}

In this paper
we applied the RG+cRPA theory to the three-orbital $d$-$p$ Hubbard model and
discovered that
the $d$-symmetry $p$-ODW
susceptibility critically develops at $\bm q=\bm Q_\mathrm{a}$
in the strong-spin-fluctuation region.
The main result is shown in 
Figs.\ \ref{fig03}(e)$-$\ref{fig03}(h).
The obtained $p$-ODW state has the characteristics:
 (i) the ordering vector is axial type
 $\bm{Q}_\mathrm{a}=(\delta_\mathrm{a},0)$,
 (ii) the charge modulation occurs on the oxygen $p$ orbital,
 and (iii) the symmetry of  the $p$-orbital order pattern is of the
 $d$-wave type.
Therefore,
the present RG+cRPA theory reproduced satisfactorily 
the experimental charge-density-wave state.
The $p$-ODW originates from the strong interference between the 
spin and orbital susceptibilities.
Such an interference is the main
characteristics
of the electronic states in underdoped cuprates.

In the previous scenarios of the spin-fluctuation-driven density-wave
states, such as  
BDW 
\cite{Metlitski:2010gf,Metlitski:2010cg,%
Holder:2012ks,
Husemann:2012eb,
Efetov:2013ib,Sachdev:2013bo,Pepin:2014tb,
Allais:2014hm,%
Chowdhury:2014cp,%
Whitsitt:2014vca}, 
PDW 
\cite{Berg:2009gt,Lee:2014ka,Fradkin:2015co,Wang:2015uw,Wang:2015iq},
and the composite orders 
\cite{Wang:2014fr,Tsvelik:2014ce},
the Maki-Thompson VC has been studied as 
possible origins 
\cite{
Metlitski:2010gf,Metlitski:2010cg,%
Holder:2012ks,%
Sachdev:2013bo,%
Chowdhury:2014cp,%
Wang:2015uw,Wang:2015iq,%
Wang:2014fr%
}.
In the weak-spin-fluctuation region, the Maki-Thompson VC is dominant and 
  the BDW instability with  $\bm q = \bm Q_\mathrm{d}$ is obtained, consistently
  with  Refs.\
\cite{
Metlitski:2010gf,Metlitski:2010cg,%
Holder:2012ks,%
Sachdev:2013bo,%
Chowdhury:2014cp,%
Husemann:2012eb,Whitsitt:2014vca%
}.
However,
in the strong-spin-fluctuation region,
 the $p$-ODW with the axial wave vector $\bm Q_\mathrm{a}$ is 
  critically enhanced:
The obtained beautiful scaling between the spin and $p$-ODW
susceptibilities means the important role of the Aslamazov-Larkin VC.
Therefore, the $p$-ODW in cuprates originates from the strong
interference between the spin and orbital fluctuations,
which is microscopically described as
the vertex corrections.

Despite that
the $p$-ODW and the $d$-orbital BDW  have the same $d$ symmetry, 
we found that
 the $p$-ODW susceptibility is strongly enhanced 
while the enhancement of the BDW susceptibility
  is moderate, 
as seen from Fig.\ \ref{fig04}(a).
This fact implies that the charge-density
modulation in cuprates is not due to the bond modulation of the $d$ orbital, but 
  due to the charge modulation on the $p$ orbital.
In fact, the recent x-ray diffraction study reported that 
the sizable oxygen-site displacements occur
in the charge-density-wave state,
whereas the Cu-site displacements are very small  \cite{Forgan:2015ud}.
This result supports the $p$-ODW scenario proposed in this paper.

The RG+cRPA theory 
developed in this study 
will be useful for 
analyzing 
unsolved problems in 
 strongly-correlated electron
systems.
For example, 
it is interesting to apply the RG+cRPA theory for 
Fe-based superconductors in order
to understand the origin of the electronic nematic state
\cite{Kasahara:2012ij,Fernandes:2010ci,Onari:2012jb,Kontani:2014ws}.

\vspace*{-.3cm}

\section*{ACKNOWLEDGMENTS}

\vspace*{-.3cm}

The authors are grateful for fruitful discussions with 
A.V. Chubukov, 
J.C.S.\ Davis,
K.\ Fujita,
T. Hanaguri, 
C. Honerkamp, 
W. Metzner,  and
S. Onari.
This work was supported by Grant-in-Aid for Scientific Research from 
the Ministry of Education, Culture, Sports, Science, and Technology, Japan.

\appendix

\begin{figure}[b]
\includegraphics[width=8cm,bb=0 0 586 270]{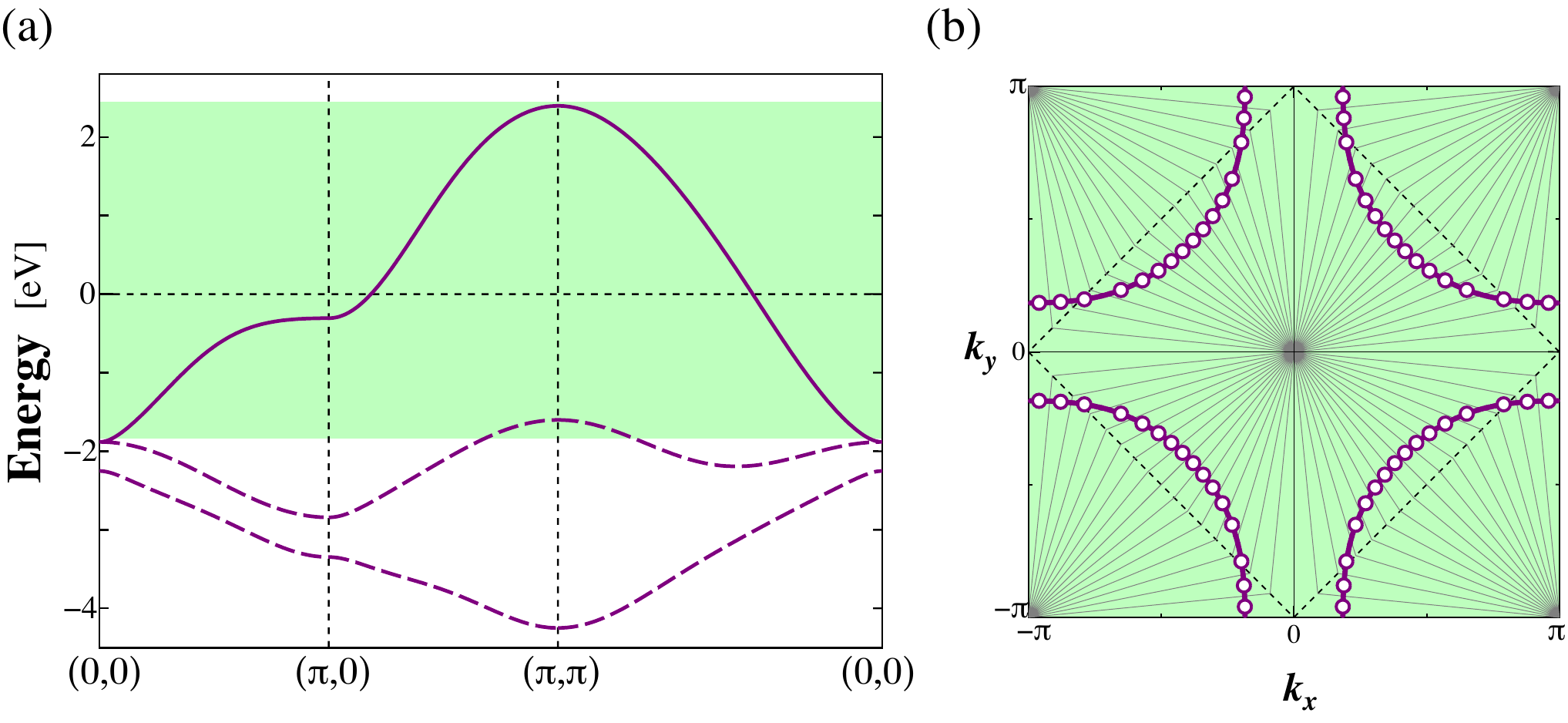}

\caption{
(a)
Band structure and (b) Fermi surface in the present model.  The
discretization of the Brillouin zone ($N_p=64$) in the conventional patch-RG
scheme are shown.
In each patch, the center momentum $\bm p(\phi_i)$ is shown as the open circle.
The initial cutoff $\Lambda_0$ is set to the bandwidth of
the conduction band, and the contributions from the valence bands are
neglected.
}
\label{fig07}
\end{figure}

\section{COMPARISON BETWEEN RG+cRPA \\ AND THE CONVENTIONAL PATCH-RG SCHEME}
\label{sec:app-A}

\vspace*{-.3cm}

In the main text,
we have analyzed the susceptibilities using the recently developed
 RG+cRPA theory \cite{Tsuchiizu:2013gu,Tsuchiizu:2015cs}.
In the conventional patch-RG method, 
the numerical accuracy for the higher-energy processes becomes worse
because of the large patch radius
 \cite{Halboth:2000vm,Halboth:2000tt,Honerkamp:2001uw,Metzner:2012jv}.
In order to improve the numerical accuracy, 
in the RG+cRPA method, 
the higher-energy processes are calculated accurately 
within the cRPA by introducing fine $\bm k$ meshes.
For this reason, the $\bm q$ dependencies of susceptibilities 
are obtained 
very accurately.
Although the effect of VCs is underestimated in the RG+cRPA method,
this underestimation is not serious since the VCs are important
mainly in the lower-energy processes.
In this section, 
we make a direct comparison between the numerical results of 
the RG+cRPA and those of 
 the conventional patch-RG method
  in order to confirm the validity and reliability
 of the 
  RG+cRPA theory.

\begin{figure}[t]
\includegraphics[width=8cm,bb=0 0 497 139]{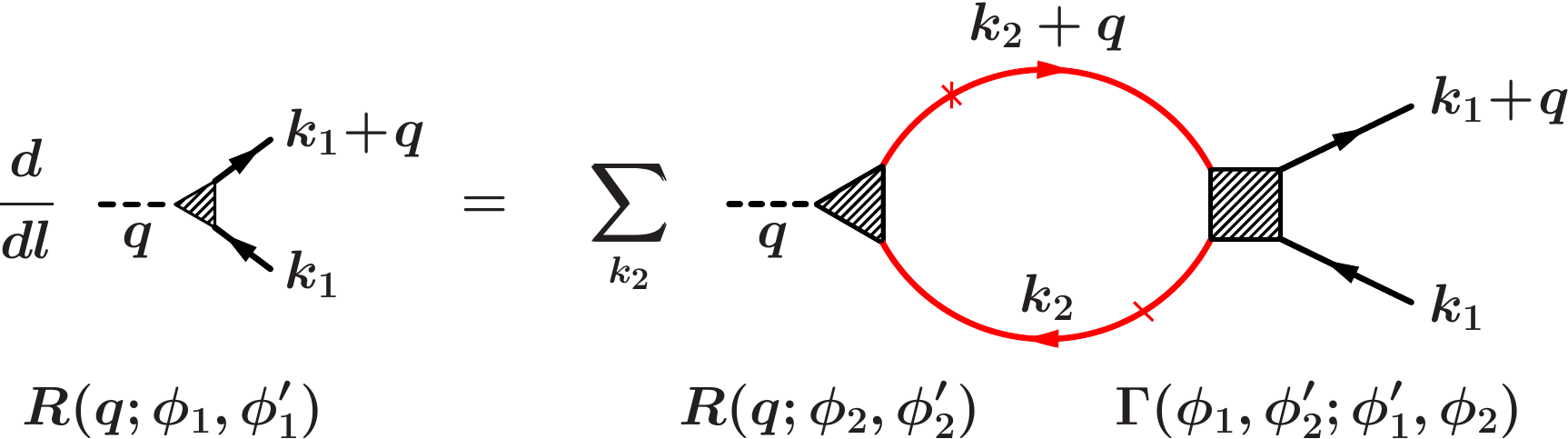}

\caption{
The RG equations for the three-point vertex $R(\bm q;\phi,\phi')$,
where $\phi$ is the patch index and $l$ is the scaling parameter.
The slashed (crossed) line  represents an electron propagation having 
  energy on $\Lambda_{l+dl}<|E|< \Lambda_l$ $(|E|>\Lambda_l)$
  where $\Lambda_l=\Lambda_0 e^{-l}$.
The patch indices are determined so that 
the momenta $\bm k_{i}$ and $\bm k_{i}+\bm q$ ($i=1,2$)
are respectively on the $\phi_{i}$ and $\phi_{i}'$ patches.
Therefore, $\bm k_i \ne \bm p(\phi_i)$ and
$\bm k_i+\bm q \ne \bm p(\phi'_i)$.
}
\label{fig08}
\end{figure}

\begin{figure}[t]
\includegraphics[width=7cm,bb=0 0 342 777]{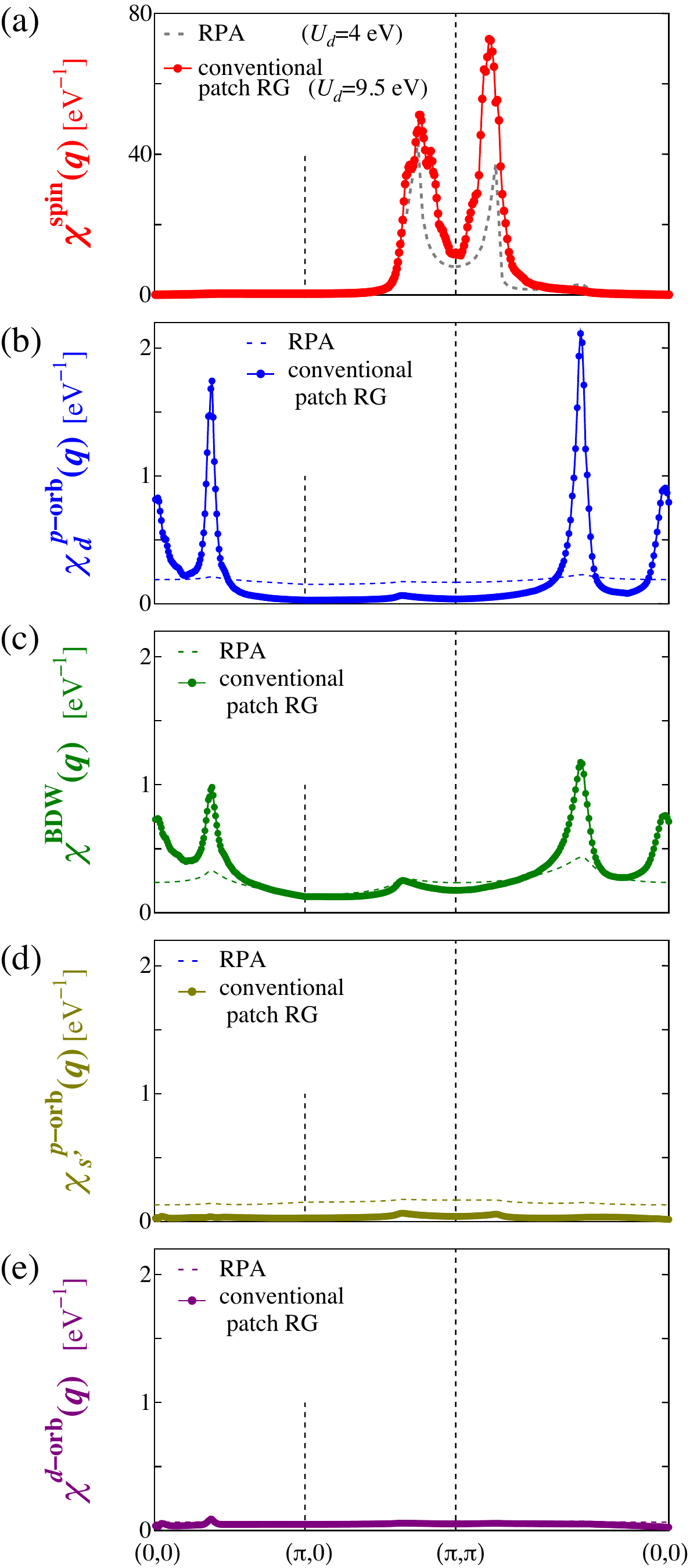}

\caption{
(a) $\chi^\mathrm{spin}(\bm q)$, 
(b) $\chi^{p\mbox{-}\mathrm{orb}}_{d} (\bm q)$, 
(c) $\chi^\mathrm{BDW} (\bm q)$, 
(d) $\chi^{p\mbox{-}\mathrm{orb}}_{s'}(\bm q)$, and
(e) $\chi^{d\mbox{-}\mathrm{orb}}(\bm q)$, 
 obtained by the conventional patch-RG method. 
We set $U_d = 9.5$ eV, $T = 20$ meV, and $N_p = 64$.
For comparison, the RPA result for $\chi^\mathrm{spin}(\bm q)$
 with $U_d = 4$ eV is also shown in (a), 
and the RPA results for $U_d=9.5$ eV are shown in (b)-(e).
As shown in (b) and (c), $d$-symmetry $p$-ODW instability 
at $\bm q=\bm Q_\mathrm{a}$ develops strongly.
These results are consistent with the results
of the RG+cRPA method explained in the main text.
}
\label{fig09}
\end{figure}

The band structure and the patch discretization 
in the conventional patch-RG method are shown in
Figs.\ \ref{fig07}(a) and \ref{fig07}(b), respectively.
The Brillouin zone is divided into the patch segments 
with respect to the angle $\phi_i$ [Fig.\ \ref{fig07}(b)].
Here, the radial variable $k_\mathrm{r}$ 
from the center of each patch, ${\bm p}(\phi_i)$,
is neglected in the vertices, that is,
$\Gamma(\{\bm k_i\})=\Gamma(\{k_{\mathrm{r},i},\phi_i\})\to \Gamma(\{\phi_i\})$
for the four-point vertex and 
$R(\bm q; \bm k_1,\bm k_2)\to R(\bm q;\phi_1,\phi_2)$
for the three-point vertex
 \cite{Halboth:2000vm,Halboth:2000tt,Honerkamp:2001uw,Metzner:2012jv}.
The justification of this approximation is frequently 
ascribed to the simple scaling argument in the weak-coupling limit.
However, 
 the  $k_\mathrm{r}$ dependencies
 are \textit{quantitatively} important 
in reality, 
since 
the ``momentum mismatch'' problem is serious
in the higher-energy processes
Moreover,
this problem becomes more serious 
 in multi-orbital 
systems, since the orbital components of each band,
called the ``orbital makeup'' \cite{Metzner:2012jv},
depend on $k_\mathrm{r}$ in usual multi-orbital models.

The conventional patch-RG scheme
can be reproduced in the present
RG+cRPA method by neglecting the valence-band contributions and 
by setting
the cutoff $\Lambda_0$ to the bandwidth of the conduction band 
  [see Fig.\ \ref{fig07}(a)].
The RG equation for $R(\bm q;\phi_1,\phi_2)$ is shown diagrammatically
 in Fig.\ \ref{fig08}.
 The susceptibilities
$\chi^\mathrm{spin}(\bm q)$, $\chi^{p\mbox{-}\mathrm{orb}}_{d}(\bm q)$, 
$\chi^\mathrm{BDW}(\bm q)$, 
$\chi^{p\mbox{-}\mathrm{orb}}_{s'}(\bm q)$,  and 
$\chi^{d\mbox{-}\mathrm{orb}}(\bm q)$
obtained by the conventional patch-RG
method for $U_d=9.5$ eV are shown in Fig.\ \ref{fig09}.
As shown in Fig.\ \ref{fig09}(a), 
strong antiferromagnetic fluctuations are obtained, although
the obtained incommensurate peak position of $\chi^\mathrm{spin}(\bm q)$ 
is at $\bm q=(\pi-\delta_\mathrm{s},\pi-\delta_\mathrm{s})$, 
differently from the experimental peak positions at 
$\bm q=(\pi,\pi-\delta_s)$ and $(\pi-\delta_s,\pi)$
that are reproduced by the RPA as well as the RG+cRPA theory 
[see Fig.\ \ref{fig02}(c)].

\begin{figure}[t]
\includegraphics[width=7cm,bb=0 0 525 383]{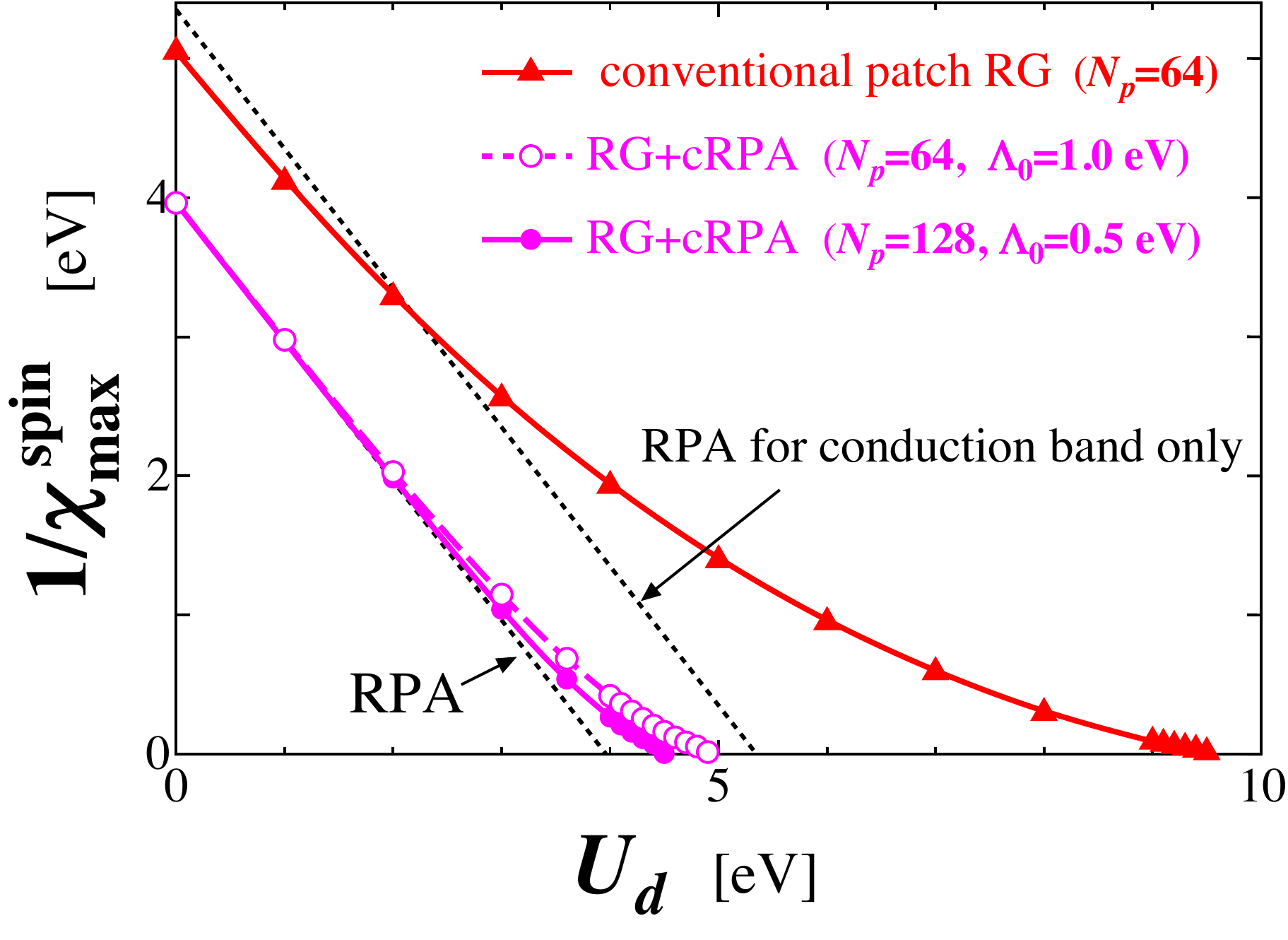}

\caption{
The $U_d$ dependence of $\chi^\mathrm{spin}_\mathrm{max}$ 
obtained by the conventional patch-RG method, RG+cRPA methods, and RPA. 
In the RG+cRPA, the results with the different choices of parameters 
are shown.
In the weak-coupling limit, the RG+cRPA results well reproduce the RPA results,
 irrespective to the choice of $N_p$.
In addition, the $\Lambda_0$ dependence of the results is weak in the RG+cRPA.
}
\label{fig10}
\end{figure}

Figure\ \ref{fig10} shows the inverse of the spin susceptibilities 
$[\chi^{\mathrm{spin}}({\bm q})]_{\rm RG+cRPA}$ and 
$[\chi^{\mathrm{spin}}({\bm q})]_\mathrm{conv\mbox{-}RG}$, 
obtained in the RG+cRPA  and  conventional patch-RG methods, respectively.
In the case of $U_d\sim0$, $[\chi^{\mathrm{spin}}({\bm q})]_{\rm RG+cRPA}$ 
almost perfectly reproduces the RPA susceptibility 
$\chi^{\mathrm{spin}}_{\rm RPA}({\bm q})= \chi^0({\bm q})/(1-U\chi^0({\bm q}))$, 
where $\chi^0({\bm q})$ 
is the bare susceptibility. 
It is noteworthy that the result of the RG+cRPA method for 
$\Lambda_0=0.5$ eV is very similar to that for $\Lambda_0=1.0$ eV.
In contrast, $[\chi^{\mathrm{spin}}({\bm q})]_\mathrm{conv\mbox{-}RG}$ 
gives under-estimated value 
since the van Vleck contribution is dropped. 
In addition, $[\chi^{\mathrm{spin}}({\bm q})]_\mathrm{conv\mbox{-}RG}$ 
also deviates from the RPA 
restricted to the conduction-band contribution, 
$\chi^{\mathrm{spin}}_{\rm RPA, cond}({\bm q})$. 
The origin of this deviation is that the orbital components,
called the ``orbital makeup''\cite{Metzner:2012jv,Maier:2013fj},
largely depend on $k_{\rm r}$ because of the large patch radius
in the conventional patch-RG method.
Thus, the RG+cRPA method is superior to the conventional patch-RG method
in the numerically accuracy for $U_d\sim0$.

Next, we compare the results of two RG methods for finite $U_d$.
As shown in Fig.\ \ref{fig10},
each $1/[\chi^{\mathrm{spin}}({\bm q})]_{\rm RG+cRPA}$ 
and $1/[\chi^{\mathrm{spin}}({\bm q})]_\mathrm{conv\mbox{-}RG}$
is concave as functions of $U_d$ because of the VC
generated in the three- and four-point vertices.
We see that $U_d^{\rm cr}\approx 4.5-5$ eV in the RG+cRPA method,
whereas $U_d^{\rm cr}\approx 9.5$ eV in the conventional patch-RG method.
To understand such large difference in $U_d^{\rm cr}$, 
in Fig.\ \ref{fig08},
we comment on the ``momentum mismatch'' between the pair of the Green functions
$G(\bm{k}_2)G(\bm{k}_2+\bm{q})$ and $\Gamma$, $R$,
originating from the fact that
the momenta at the centers of the patches $\phi_2$ and $\phi_2'$, 
${\bm p}(\phi_2)$ and ${\bm p}(\phi_2')$,
are different from $\bm{k}_2$ and $\bm{k}_2+\bm{q}$, respectively.
This mismatch leads to violation of the momentum conservation
at each bare vertex $U_d$ in the Feynman diagrams.
This defect is not serious in the RG+cRPA method with $\Lambda_0\ll W_{\rm band}$.
Considering that the self-energy correction is dropped 
in the present RG methods,
large $U_d^{\rm cr}$ in the conventional patch-RG method
might partially originate from the momentum mismatch.
On the other hand, $U_d^{\rm cr}$ in the RG+cRPA method
is underestimated since the vertex corrections 
with higher-energy processes ($|E|>\Lambda_0$)
are included only partially.
In spite of this underestimation,
large $p$-ODW susceptibility is induced by the VC in the RG+cRPA method.

We stress that
the results of the RG+cRPA method and those 
of the conventional patch-RG method are qualitatively equivalent,
nonetheless of the large difference in $U_d^{\rm cr}$.
In the RG+cRPA method,
the results are almost unchanged
in the both cases of $\Lambda_0=0.5$ eV and $\Lambda_0=1.0$ eV.
In addition, 
the RG+cRPA results with $N_p=64$ and $128$ are almost identical,
 indicating that the 
the numerical convergence has been achieved at  $N_p\approx 64$.
We note that various RG methods have been proposed to resolve the 
above-mentioned problems in the conventional patch-RG method,
such as the singular-mode functional renormalization group
\cite{Maier:2013fj}.

Regarding the charge susceptibilities,
both $\chi^{p\mbox{-}\mathrm{orb}}_{d}(\bm q)$
 and $\chi^\mathrm{BDW}(\bm q)$ possess large sharp peaks 
at $\bm q=\bm Q_\mathrm{a}$ and $\bm Q_\mathrm{d}$ 
as shown in Figs.\ \ref{fig09}(b) and \ref{fig09}(c) at $U_d=9.5$ eV.
This behavior is essentially identical to that obtained 
 by RG+cRPA method  (Fig.\ \ref{fig03}).
This result supports the validity of the present RG+cRPA theory.
Since the van Vleck contributions are dropped 
in the conventional patch-RG method,
both $\chi^{p\mbox{-}\mathrm{orb}}_{d}(\bm q)$ 
and $\chi^{p\mbox{-}\mathrm{orb}}_{s'}(\bm q)$ 
in Figs.\ \ref{fig09}(b) and \ref{fig09}(d) are much smaller than the RPA results
except near the peak positions.
In the RG+cRPA, in contrast,
    the RG+cRPA results coincide well with the RPA results 
if $\bm q$ is away from the 
peak positions, as shown in Fig.\ \ref{fig03}.

\begin{figure}[t]
\includegraphics[width=7cm,bb=0 0 562 462]{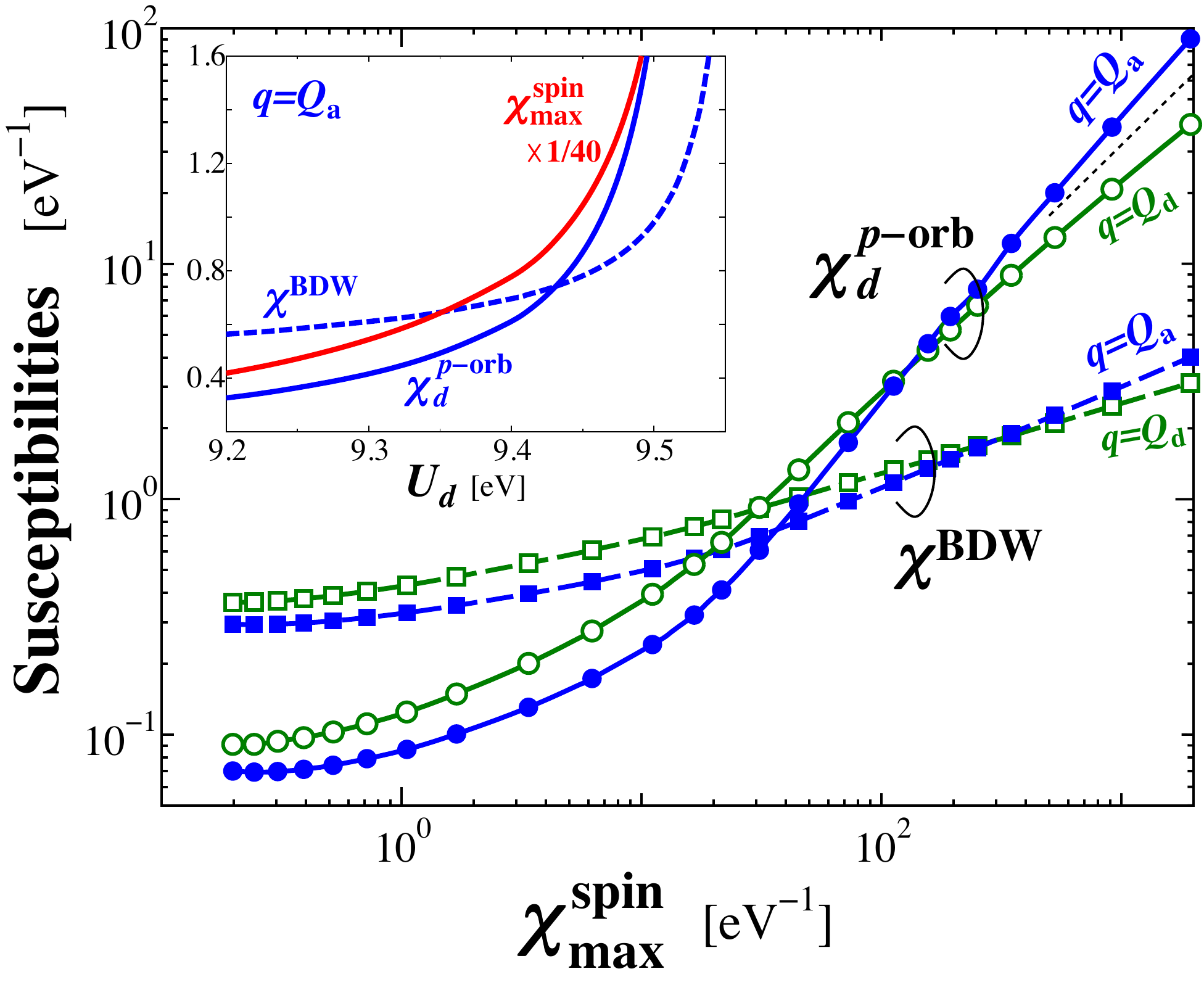}
\caption{
Peak values of $\chi^{p\mbox{-}\mathrm{orb}}_{d}(\bm q)$ 
and $\chi^\mathrm{BDW}(\bm q)$ 
as functions of $\chi^\mathrm{spin}_\mathrm{max}$ 
obtained from the conventional patch-RG method.  
The gradient of the dotted line is 1, so
beautiful scaling relation 
$\chi^{p\mbox{-}\mathrm{orb}}_{d}(\bm q) \propto 
 \chi^\mathrm{spin}_\mathrm{max}$ is satisfied
in the strong-coupling regime.
The inset shows the $U_d$ dependencies of 
$\chi^{p\mbox{-}\mathrm{orb}}_{d}(\bm q)$ and $\chi^\mathrm{BDW}(\bm q)$ 
at $\bm q=\bm Q_\mathrm{a}$. 
}
\label{fig11}
\end{figure}

Figure \ref{fig11} shows the scaling relation between 
the $p$-ODW susceptibility and
$\chi^\mathrm{spin}_\mathrm{max}$ obtained 
from the conventional patch-RG method.  The obtained
relations $\chi^{p\mbox{-}\mathrm{orb}}_{d}(\bm Q_\mathrm{a})
  \propto \chi^\mathrm{spin}_\mathrm{max}$ as well as 
$\chi^{p\mbox{-}\mathrm{orb}}_{d}(\bm Q_\mathrm{a}) \gg 
\chi^\mathrm{BDW}(\bm Q_\mathrm{a})$ 
in the strong-spin-fluctuation region are
qualitatively consistent with the RG+cRPA results shown in 
Fig.\ \ref{fig05}.
 Therefore, the development of the $d$-symmetry axial $p$-ODW
instability due to the spin fluctuation, which is the main
message of the main text derived from the RG+cRPA, is confirmed also by the
conventional patch-RG method.

In summary, the results given by the RG+cRPA method presented in the
main text are qualitatively reproduced by the conventional patch-RG
method.  Therefore, the validity of the RG+cRPA method is verified by
the conventional patch-RG 
method, the validity of which has been confirmed
in literature 
 \cite{Halboth:2000vm,Halboth:2000tt,Honerkamp:2001uw,Metzner:2012jv}.
The numerical accuracy is well improved in the RG+cRPA method, as
recognized by the coincidence with the RPA result in the weak-coupling
region shown in Fig.\ \ref{fig10}.
This improvement is achieved by
calculating the higher-energy processes accurately within the cRPA
by introducing the fine $\bm k$ meshes.
As seen from 
Figs.\ \ref{fig02}(c) and \ref{fig02}(d), 
the contributions from the cRPA are small
for $\Lambda_0=0.5$ eV but quite important 
(especially for the four-point vertex
 \cite{Tsuchiizu:2013gu,Tsuchiizu:2015cs})
 in order to derive reliable results from the RG equations.

\begin{figure}[t]
\includegraphics[width=7cm,bb=0 0 478 295]{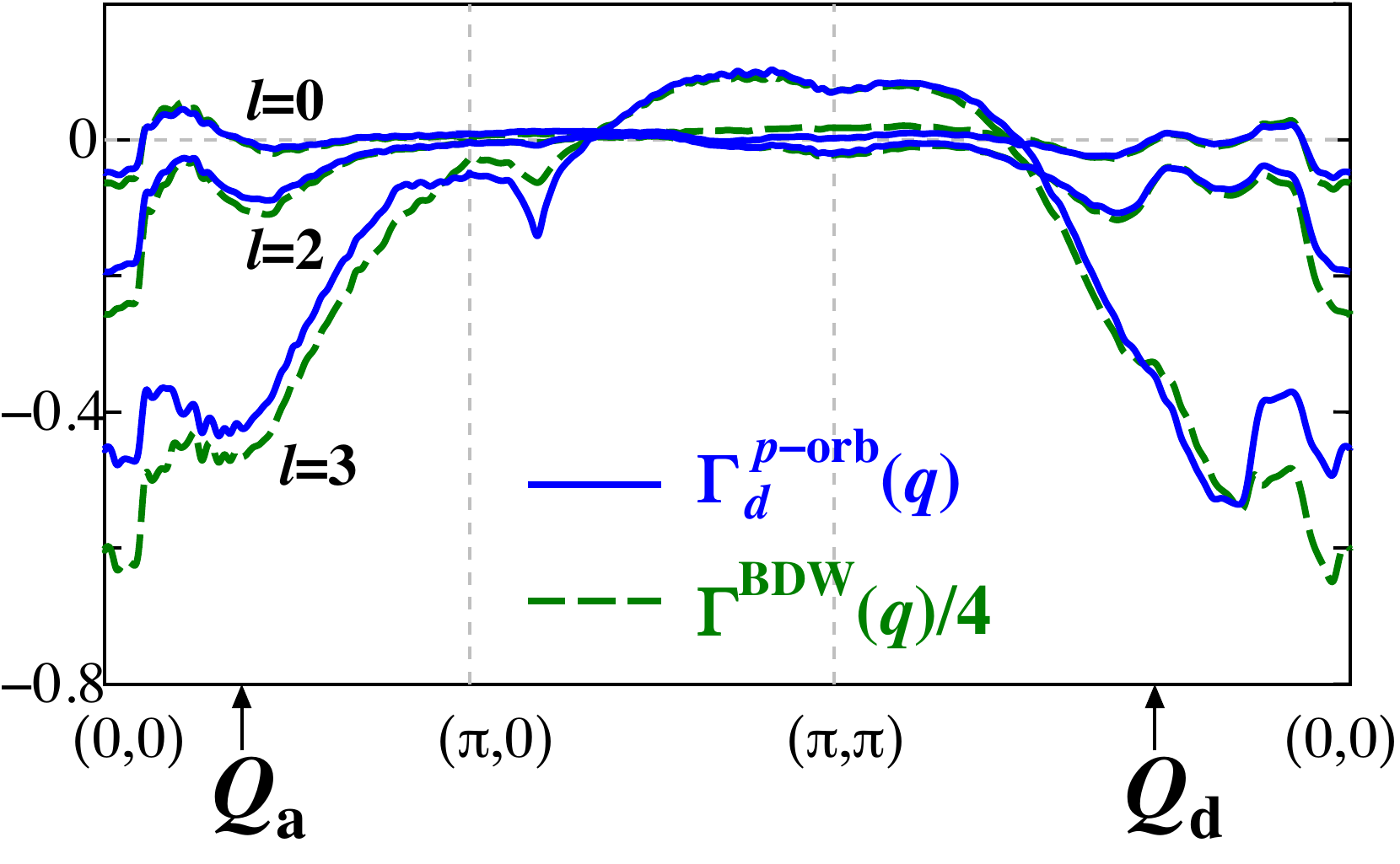}
\caption{
The effective interactions 
(\ref{eq:effint-BDW})
and 
(\ref{eq:effint-pODW})
at the scaling parameter $l=0$, $2$, and $3$, 
with $U_d=4.50$ eV, $T=20$ meV, $N_p=128$.
In this case,  $\chi^{p\mbox{-}\mathrm{orb}}_d(\bm Q_\mathrm{a})$
is larger than $\chi^{\mathrm{BDW}}(\bm Q_\mathrm{d})$
 [see Fig.\ \ref{fig03}(h)].
}
\label{fig12}
\end{figure}

\begin{figure*}[t]
\includegraphics[width=15cm,bb=0 0 1458 274]{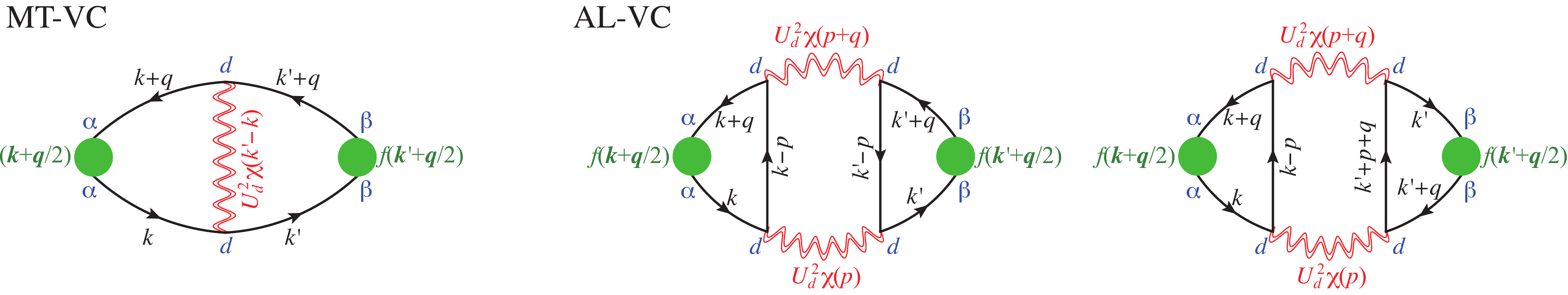}
\caption{
Detailed diagrammatic expressions 
of the Maki-Thompson and Aslamazov-Larkin VCs.
Their analytic equations are shown in Eqs.\
(\ref{eq:SM-VC1})$-$(\ref{eq:SM-VC4}).
}
\label{fig13}
\end{figure*}

\vspace*{-.3cm}

\section{SCALING FLOWS OF EFFECTIVE INTERACTIONS}
\label{sec:app-B}

\vspace*{-.3cm}

In this section,
we examine the effective interactions for $p$-ODW and $d$-orbital BDW.
The effective interaction for the $d$-orbital BDW state is 
  given by taking into account the $d$-form factor
\begin{eqnarray}
\Gamma^\mathrm{BDW}(\bm q)
&\equiv&
{\sum_{\bm k ,\bm k'}}'
\Gamma^c (\bm k,\bm k' ; \bm k+\bm q , \bm k'-\bm q)
\nonumber \\ && {} \times
u_{d}^*(\bm k) \, 
u_{d}^*(\bm k') \, 
u_{d}(\bm k+\bm q) \, 
u_{d}(\bm k'-\bm q) 
\nonumber \\ && {} \times
f(\bm k + \bm q /2) \, f(\bm k' - \bm q /2),
\label{eq:effint-BDW}
\end{eqnarray}
where $\Gamma^c(\bm k ,\bm k'; \bm k+\bm q, \bm k'-\bm q)$ represents 
the renormalized charge vertex.
This expression is essentially the same as examined in the 
 single-orbital case \cite{Husemann:2012eb}.
Here, $u_d(\bm k)$ represents  the unitary transformation 
 from the conduction band basis to the $d$-orbital basis.
The summations over $\bm k$ and $\bm k'$ are  restricted to the
low-energy scattering processes,
in which 
all the momenta
$(\bm k,\bm k' , \bm k+\bm q , \bm k'-\bm q)$ 
are near the 
Fermi surface.
This interaction becomes zero 
if the bare Hubbard interaction $U_d$ is used for
$\Gamma^c$  \cite{Husemann:2012eb}.
In the present case, $\Gamma^{\mathrm{BDW}}|_{l=0}$ becomes
small but nonzero due to the cRPA contributions.

In the similar manner, 
the effective interaction
 for the $p$-ODW state can be introduced as
\begin{eqnarray}
\Gamma_d^{p\mbox{-}\mathrm{orb}}(\bm q)
&\equiv&
\Gamma^{c}_{xxxx}(\bm q)
+
\Gamma^{c}_{yyyy}(\bm q)
\nonumber \\ && {}
-
\Gamma^{c}_{xyxy}(\bm q)
-
\Gamma^{c}_{yxyx}(\bm q),
\label{eq:effint-pODW}
\\
\Gamma^{c}_{\alpha\beta\gamma\delta}(\bm q)
&\equiv&
{\sum_{\bm k ,\bm k'}}'
\Gamma^c (\bm k,\bm k' ; \bm k+\bm q , \bm k'-\bm q)
\nonumber \\ && {} \times
u_{\alpha}^*(\bm k) \, 
u_{\beta}^*(\bm k') \, 
u_{\gamma}(\bm k+\bm q) \, 
u_{\delta}(\bm k'-\bm q) ,
\qquad
\end{eqnarray}
where $u_\alpha(\bm k)$ ($\alpha=x,y$) 
represents  the unitary transformation 
 from the band basis to the $p_x$- or $p_y$-orbital basis.

As has been discussed in Ref.\ \cite{Husemann:2012eb}, 
  the negative values of the effective interactions
indicate the precursor 
  of the corresponding instability.
Actually, if the effective interaction becomes negative,
the rhs of the RG equations for 
 the  three-point vertex 
\cite{Tsuchiizu:2013gu} becomes positive 
 and then the three-point vertex can  increase.
Once the three-point vertex is enhanced, the corresponding
susceptibility  is also enhanced \cite{Tsuchiizu:2013gu}.

In Fig.\ \ref{fig12}, we show the development of the effective interactions 
for several scaling parameters $l=\ln(\Lambda_0/\Lambda)$
in the case of $U=4.50$ eV.
At the initial point ($l=0$), these effective interactions
given by the cRPA are small and almost independent of $\bm q$.
After the renormalization ($l=2,3$), the effective interactions at 
$\bm Q_\mathrm{a}$ and $\bm Q_\mathrm{d}$ become negative,
implying the development of the susceptibilities.
At $l=3$, the interaction at $\bm q=\bm Q_{\mathrm{a}}$ is
  stronger than that at $\bm q=\bm Q_{\mathrm{d}}$.
We find that the overall profile of $\Gamma^{\mathrm{BDW}}(\bm
  Q_\mathrm{a})$
is similar to $\Gamma^{p\mbox{-}\mathrm{orb}}(\bm Q_\mathrm{a})$,
 except for the constant factor.

In contrast, in the weak-fluctuation case ($U_d=4.40$ eV),
 the interaction at $\bm q=\bm Q_{\mathrm{a}}$ becomes weaker than 
that at $\bm q=\bm Q_{\mathrm{d}}$.
This relation of the effective interactions 
is in accordance with that of the peak positions 
  of the susceptibility.

Thus, we understand that the enhancements of $\chi^\mathrm{BDW}(\bm q)$ 
and $\chi^{p\mbox{-}\mathrm{orb}}_d(\bm q)$ at $\bm q=\bm Q_\mathrm{a}$,
$\bm Q_\mathrm{d}$, and $\bm 0$ originate from the large negative 
$\Gamma^\mathrm{BDW}(\bm q)$ and $\Gamma^{p\mbox{-}\mathrm{orb}}(\bm q)$
after the renormalization.
However, the $\bm q$ dependencies of 
$\Gamma^{\mathrm{BDW}}(\bm q)$ and
$\Gamma^{p\mbox{-}\mathrm{orb}}(\bm q)$
are not consistent with those of the susceptibilities.
For example, the peak position of each 
$\chi^{\mathrm{BDW}}(\bm q)$ and
$\chi^{p\mbox{-}\mathrm{orb}}_d(\bm q)$ is
 $\bm q=\bm Q_\mathrm{a}$ or $\bm Q_\mathrm{d}$ for $U_d=4.50$ eV
as shown in 
Figs.\ \ref{fig03}(e)$-$\ref{fig03}(h), although both
$|\Gamma^{\mathrm{BDW}}(\bm q)|$ and $|\Gamma^{p\mbox{-}\mathrm{orb}}(\bm q)|$ 
take the maximum values at $\bm q=\bm 0$.
These results mean that the susceptibilities should be calculated to
obtain the real density-wave wave vector in the RG study.

\vspace*{-.3cm}

\section{ANALYTICAL EXPRESSIONS OF THE MAKI-THOMPSON AND
ASLAMAZOV-LARKIN VERTEX CORRECTIONS}
\label{sec:app-C}

\vspace*{-.3cm}

In this section, we show the analytical expressions of the leading-order
Aslamazov-Larkin VC $X^{{\rm AL} c}_{l, m} (q)$ and Maki-Thompson VC $X^{{\rm MT} c}_{l, m} (q)$
for the charge channel, including 
the form factor $f_\alpha ( \bm{k} )$. 
The Maki-Thompson VC is given by
\begin{eqnarray}
X^{{\rm MT} c}_{\alpha, \beta} (q)
	&=&	T^2 U_{d}^2
		\sum_{k, k'}
		f_{\alpha} \Big( \bm{k} + \frac{\bm{q}}{2} \Big) \, 
		G_{\alpha, d} ( k + q ) \,
		G_{d, \alpha} ( k )
\nonumber \\ && {} \times
\left[
		\frac{3}{2}
		\chi^\mathrm{spin} ( k - k' )
+\frac{1}{2}
		\chi^\mathrm{charge} ( k - k' )
\right]
\nonumber \\ && {} \times
		G_{\beta, d} ( k' ) \,
		G_{d, \beta} ( k' + q ) \,
		f_{\beta} \Big( \bm{k}' + \frac{\bm{q}}{2} \Big)
,
\end{eqnarray}
where $k = ( \bm{k}, \epsilon_{l})$, 
$q = ( \bm{q}, i \omega_{m})$,
$ \epsilon_{l} = \pi T (2 l + 1) $, 
$ \omega_{m} = 2 \pi m T$,
and $\alpha$, $\beta$ $= d, x, y$ represent the orbital indices. 
Here, $\chi^\mathrm{spin}$ 
 and $\chi^\mathrm{charge}(\equiv \chi^{d\mbox{-}\mathrm{orb}})$ 
represent the $d$-orbital spin and charge susceptibilities,
given by Eqs.\ (\ref{eq:d-orb-spin})  and (\ref{eq:d-orb-charge}), respectively. 
The Aslamazov-Larkin VC is given by
\begin{eqnarray}
X^{{\rm AL} c}_{\alpha, \beta} (q)
	&=&	T U_{d}^4
		\sum_{p}
		\Lambda_{\alpha} ( q ; p )
\biggl[
\frac{3}{2}
		\chi^\mathrm{spin} ( p + q )
		\chi^\mathrm{spin} ( p )
\nonumber \\ && {} +
		\frac{1}{2}
		\chi^\mathrm{charge} ( p + q )
		\chi^\mathrm{charge} ( p )
\biggr]
		\Lambda_{\beta}' ( q ; p )
,
\end{eqnarray}
where 
$p = ( \bm{p}, i \omega_{n})$ and $ \omega_{n} = 2 \pi n T$.
Here $\Lambda_{l}$ and $\Lambda_{l}'$ are the three-point vertices:
\begin{eqnarray}
\Lambda_{\alpha} ( q ; p )
	&=&	T
		\sum_{k}
		f_{\alpha} \Big( \bm{k} + \frac{\bm{q}}{2} \Big) \,
\nonumber \\ && {} \times
		G_{\alpha, d} ( k + q ) \,
		G_{d, d} ( k - p ) \,
		G_{d, \alpha} ( k )
,
\label{eq:SM-VC1}
\\
\Lambda_{\beta}' ( q; p )
	&=&	T
		\sum_{k'}
		f_{\beta} \Big( \bm{k}' + \frac{\bm{q}}{2} \Big)
		G_{\beta, d} ( k' )
		\bigl[
			G_{d,d} ( k' - p )
\nonumber \\ && {} 
		+	G_{d,d} ( k' + p + q )
		\bigr]
		G_{d, \beta} ( k' + q )
.
\end{eqnarray}
Their diagrammatic expressions are shown in Fig.\ \ref{fig13}.

For the $d$-orbital BDW with $d$-symmetry form factor, 
the form factors $f_\alpha(\bm k)$ ($\alpha=d,x,y$) are set to 
\begin{equation}
f_{d} (\bm{k} )
= \cos ( k_{x} ) - \cos ( k_{y} ) 
,\quad
f_{x} (\bm{k} ) =
f_{x} (\bm{k} ) = 0. 
\end{equation}
The form factors for the $p$-ODW with $d$ symmetry  are
\begin{eqnarray}
f_{d} (\bm{k} ) = 0,\quad
f_{x} (\bm{k} )= +1 , \quad
f_{y} (\bm{k} )= -1,
\label{eq:SM-VC4}
\end{eqnarray}
and those with $s'$ symmetry are
\begin{eqnarray}
f_{d} (\bm{k} ) = 0,\quad
f_{x} (\bm{k} )= +1 , \quad
f_{y} (\bm{k} )= +1 .
\end{eqnarray}



\end{document}